\documentclass[fleqn,usenatbib]{mnras}
\usepackage{newtxtext,newtxmath}

\usepackage[T1]{fontenc}

\DeclareRobustCommand{\VAN}[3]{#2}
\let\VANthebibliography\thebibliography
\def\thebibliography{\DeclareRobustCommand{\VAN}[3]{##3}\VANthebibliography}

\usepackage{graphicx}	
\usepackage{amsmath}	

\usepackage{amssymb}	
\usepackage{color}      
\usepackage{array}
\usepackage[flushleft]{threeparttable}
\usepackage{booktabs,caption,soul,multirow}

\newcommand{\Oabu}{12+\log(\mathrm{O/H})}

\newcommand{\hii}{\mathrm{H~\textsc{ii}}}

\newcommand{\aref}[1]{\hyperref[#1]{Appendix~\ref{#1}}}
\newcommand{\nebbayes}{\textsc{NebulaBayes}}
\newcommand{\mappings}{\textsc{Mappings}}

\title[Strategies for future metallicity measurement]{Optimal strategies for measuring gas-phase metallicities in intermediate-redshift non-AGN and AGN-host galaxies using future instrumentation}

\author[S.-L. Li et al.]{Song-Lin Li,$^{1,2}$\thanks{E-mail: songlin.li@anu.edu.au}
Trevor J. Mendel,$^{1, 2}$
Mark R. Krumholz,$^{1, 2}$
Emily Wisnioski$^{1, 2}$
\\
$^{1}$Research School of Astronomy \& Astrophysics, Australian National University, Canberra, ACT 2611, Australia\\
$^{2}$ARC Centre of Excellence for All-Sky Astrophysics in 3 Dimensions (ASTRO 3D), Canberra, ACT 2611, Australia\\
}

\date{Accepted XXX. Received YYY; in original form ZZZ}

\pubyear{2025}

\begin{document}
\label{firstpage}
\pagerange{\pageref{firstpage}--\pageref{lastpage}}
\maketitle

\begin{abstract}
Resolved measurements of gas-phase metallicities in galaxies that both do and do not host active galactic nuclei (AGN) are crucial for understanding the role of AGN in regulating galaxy growth over cosmic time. Recent work at $z=0$ has demonstrated that a self-consistent comparison of metallicities in AGN-host and non-AGN galaxies can be conducted within a Bayesian framework given sufficient coverage of rest-frame optical emission lines. The next generation of adaptive optics-assisted optical and near-infrared integral field spectrographs promise to deliver the improved sensitivity and spatial resolution required conduct comparable measurements at intermediate redshift ($z\sim 1-3$), albeit with a restricted set of emission lines dictated by the various filter and grating combinations available. In this work, we explore optimal strategies for recovering the metallicities of both AGN-host and non-AGN galaxies given these constraints. We consider suites of emission lines that will be practical to obtain at different redshifts, and test a range of strategies to measure metallicity using just these lines, evaluating their performance by comparison to measurements using all the lines to which we have access in $z=0$ optical spectra. Our results facilitate straightforward estimates of exposure times required by future instruments to reach specified accuracy goals in a range of redshift windows using optimal sets of lines, enabling the development of efficient observing strategies for future surveys. This study can be extended to spatially resolved galaxies in order to design optimal strategies for measuring metallicity fluctuation maps in addition to the mean metallicities we consider here.
\end{abstract}




\section{Introduction}
\label{sec:intro}

Gas-phase metals record the interplay between star formation activity, gas inflow, and gas outflow during galaxy assembly, and thus gas-phase metallicity is one of the key parameters that describes galaxy formation and evolution \citep[e.g.,][and references therein]{maiolino2019}. Because it is the most abundant element in the universe after hydrogen and helium, oxygen, the abundance of which is traditionally expressed as $\Oabu$, is usually used as a proxy for this quantity. Metallicity can be estimated for $\hii$ regions from emission-line intensity ratios of species that are sensitive to metal abundances. The direct method, which is generally the most accurate approach, measures the electron temperature by the ratio of strong emission lines to auroral lines, typically [O~\textsc{iii}]$~\lambda 4363$, and estimates the metallicity from the electron temperature, which is sensitive to metallicity because metals are the dominant coolants in $\hii$ regions \citep{andrews2013,perez2017}. However, due to the weakness of the [O~\textsc{iii}]$~\lambda 4363$ emission line, especially in metal rich systems, this method is usually inapplicable to galaxies with $\Oabu > 8.7$ \citep{stasinska2005}. The most common fallback approach in this case is to develop metallicity calibrations based on the ratios of stronger emission lines. One determines the mapping from strong line ratio to metallicity either empirically by calibrating against metallicities measured by the direct method where possible \citep[e.g.,][]{pettini2004,yin2007,perez2009,brown2016}, or theoretically based on calculations using photoionisation models such as \mappings~\citep{sutherland2017} or \textsc{cloudy} \citep{Chatzikos2023}. The latter approach has the advantage that it allows the development of strong-line calibrators over a very large range of the parameter space \citep[e.g.,][]{kewley2002,kobulnicky2004,dors2011}. However, the method can introduce significant systematic uncertainties as different strong-line diagnostics show unique dependence on gas pressure and ionization parameters \citep[and references therein]{kewley2019b}. 

Most metallicity diagnostics have been developed for $\hii$ regions, requiring that galaxies that host active galactic nuclei (AGN) be removed from study samples. However, some authors have attempted to measure the metallicities for AGN hosts by deriving diagnostics for AGN-like ionizing spectra \citep[e.g.,][]{carvalho2020,dors2020,dors2021}. Because these diagnostics are derived based on pure AGN ionizing spectra, however, they cannot be applied to galaxies with significant contributions to their spectra from $\hii$ regions, precluding consistent comparisons of AGN-host galaxies and star-forming (SF) galaxies that do not host AGN. One way around this problem is by exploiting the spatial resolution offered by integral field spectroscopy (IFS). In IFS observations, the spaxels within galaxies that host AGN form a clear transitional sequence on the Baldwin, Phillips \& Terlevich (BPT) diagram \citep{baldwin1981} from the so-called BPT-SF region (where the dominant ionizing source is massive stars) to the AGN region (where the dominant ionizing source is AGN), reflecting the relative contributions of AGN and stellar photoionization to the final emission line flux. In such a map one can scale the relative fractions of strong-line fluxes from pure $\hii$ regions of individual spaxels based on their positions on BPT diagram, and extract pure $\hii$-region fluxes to measure metallicity \citep[e.g.,][]{davies2014,davies2016,dagnostino2018,dagostino2019a,dagostino2019b}.

An alternative way to disentangle SF and AGN mixing is by using a Bayesian method. \citet{thomas2018} developed the code \nebbayes~ to estimate galactic properties including metallicity under the assumption that the total spectrum represents a linear combination of emission from $\hii$ regions and narrow-line regions (NLRs), each of which is characterised by a set of photoionisation model grids derived using the appropriate driving ionizing spectrum. With this method one can measure the metallicity in galaxies ranging from pure SF galaxies to Seyfert galaxies in a self-consistent manner, enabling direct comparison between AGN-host and non-host galaxies -- a key measurement if we are to understand the role of AGN in regulating galaxy evolution. \citet{thomas2019} use \nebbayes~ to investigate the mass-metallicity relations of Seyfert galaxies in the Sloan Digital Sky Survey (SDSS), and \citet{li2024} extend this work to fundamental metallicity relations (FZR) of AGN-host and non-AGN galaxies drawn from the MaNGA \citep[Mapping Nearby Galaxies at Apache Point Observatory][]{bundy2015} survey. They find that AGN-host galaxies generally have higher metallicity than non-AGN galaxies when controlling the stellar mass due to AGN-hosts' systematically lower star formation rates (SFR), i.e., by the existence of the FZR.

With IFS surveys it is also possible to explore the abundance patterns in 2D metalicity fields including metallicity gradients \citep[e.g.,][]{sanchez2014,ho2015,belfiore2017,sharda2021b} and correlation statistics \citep[e.g.,][]{Kreckel19,Kreckel20,Metha21,Metha22,liz2021,liz2023}. Again, most authors remove AGN spaxels or galaxies hosting AGN for these analyses due to difficulties in calibrating AGN and SF metallicities simultaneously. However, \citet{li2025} extend this work by comparing the correlation lengths that characterise the metal fields of AGN-host galaxies and non-AGN galaxies in CALIFA using \nebbayes; they find that AGN only indirectly influence the metallicity fields by reducing star-forming activity, but that once one controls for (SFR) there is no remaining systematic difference between AGN-host and non-AGN galaxies, suggesting that the SFR is more fundamental in regulating metallicity fields than stellar mass and the size.

To date all of the work we have described comparing AGN-host and non-host galaxies has focused on local, $z \lesssim 0.1$, galaxies. However, for the purposes of understanding the role of AGN in galaxy evolution and formation, it is highly desirable to extend these comparisons to higher redshift, where AGN activity is both more prevalent and stronger \citep{shankar2009}.

However, such measurements are not possible with existing IFS systems, which lack both the light-gathering capability and resolution to measure metal fields in more distant targets. However, the next generation of IFS instruments that will be integrated with adaptive optics (AO) facilities such as MAVIS \citep[Multi-conjugate Adaptive-optics Visible Imager-Spectrograph;][]{rigaut2020,rigaut2021}, GMTIFS \citep[Giant Magellan Telescope Integral-Field Spectrograph;][]{mcgregor2012}, and HARMONI \citep[High Angular Resolution Monolithic Optical and Near-infrared Integral field spectrograph;][]{thatte2021} will be able to push the angular resolution to the diffraction limit on 8 m and larger telescopes, and are therefore ideal for this purpose. However, due to both the configuration of these instruments and the existence of strong atmospheric absorption features in the near infrared, only a subset of optical emission lines at $z=0$ currently used to study the difference between AGN and non-AGN host metallicities can be observed simultaneously. For example, the emission line ratio [N~\textsc{ii}]$~\lambda 6583/$[O~\textsc{ii}]$~\lambda\lambda 3727,29$, proven to be the most reliable metallicty diagnostic usable in both AGN hosts and SF galaxies \citep{kewley2002,zhang2017,kewley2019b}, will not be available in observations with these instruments, since the required lines are located at the two extreme ends of the optical band. 

Given these constraints, the best observing strategy remains unclear. Ideally we would like to be able to compare AGN-host and non-AGN galaxies at intermediate redshift, $z\sim 1 - 3$, both to each other and to local, $z\sim 0$, galaxies, with as few systematic errors as possible, and we would like to do so without requiring expensive repeated observations of a single target in order to capture lines that fall into different bands. This work presents different strategies for making use of the line ratios in emission line groups, for the purpose of measuring the metallicity in both AGN hosts and non-AGN galaxies. In \autoref{sec:ins_z}, we briefly introduce the three instruments on which we focus in this work, and summarise the suites of emission lines accessible to them for galaxies in different redshift windows. In \autoref{sec:met_cali}, we introduce the local galaxy sample that we will use for our experiment and describe our analysis method. We then present the results of our tests of different observing strategies for that sample in \autoref{sec:analysis}. We extend our results to different SNRs in \autoref{sec:snr}, creating a reference for readers for who wish to calculate exposure times required to reach targeted uncertainty levels in future surveys. We finally summarise our results in \autoref{sec:summary}.

Throughout the work, we adopt a \cite{chabrier2003} initial mass function (IMF) and a flat WMAP7 cosmology: $H_0=70.4 \,\mathrm{km\; s^{-1} \,Mpc^{-1}}$, $\Omega_M=0.27$, and $\Omega_{\Lambda}=0.73$ \citep{komatsu2011}.

\begin{table*} 
 \centering
 \renewcommand{\arraystretch}{1.3}
 \begin{threeparttable}
  \caption{Wavelength ranges of the spectral bands available for MAVIS, GMTIFS, and HARMONI that we consider in this work (left three columns), together with the redshift ranges for which these spectral bands include all of the lines in the emission line suites that we consider: the Blue, LRed, and SRed suites (right three columns); see \autoref{sec:z_range} for details.}
  \vspace{1mm}
  \label{tab:redshift}

  \begin{tabular}{lllccc}
   \hline
   \multicolumn{3}{c}{Observing setup} & \multicolumn{3}{c}{Redshift windows for line suites} \\
   Instrument & Band & Wavelength (nm) & Blue & LRed & SRed \\
   \hline
   \multirow{2}{*}{MAVIS} & LR-Blue & $\phantom{0}370 - \phantom{0}720$ & $0\phantom{.000} - 0.438$ & $0\phantom{.000} - 0.070$ & $0\phantom{.000} - 0.070$ \\
                          & LR-Red & $\phantom{0}510 - 1000$ & $0.368 - 0.997$ & $0.049 - 0.486$ & $0\phantom{.000} - 0.486$ \\
   \hline
   \multirow{4.6}{*}{GMTIFS} & $mZJ$ & $\phantom{0}890 - 1350$ & $1.387 - 1.696$ & $0.831 - 1.006$ & $0.413 - 1.006$ \\
                           & \multirow{2}{*}{$mJH$$^\mathrm{a}$} & $1190 - 1350$ & $2.192 - 2.489$ & $1.448 - 1.674$ & $0.889 - 1.006$ \\[-1ex]
                           & &  $1420 - 1800$ & -- & -- & $1.254 - 1.674$ \\
                           & \multirow{2}{*}{$mHK$$^\mathrm{b}$} & $1640 - 1800$ & $3.399 - 3.652$ & $2.374 - 2.595$ & $1.603 - 1.674$ \\[-1ex]
                           & & $1940 - 2490$ & -- & -- & $2.079 - 2.699$ \\
   \hline
   \multirow{3.8}{*}{HARMONI} & $V+R$ & $\phantom{0}458 - \phantom{0}820$ & $0.229 - 0.638$ & $0\phantom{.000} - 0.218$ & $0\phantom{.000} - 0.218$ \\
                            & $Iz+J$$^\mathrm{c}$ & $\phantom{0}811 - 1350$ & $1.175 - 1.696$ & $0.668 - 1.006$ & $0.287 - 1.006$ \\
                            & \multirow{2}{*}{$H+K$$^\mathrm{b}$} & $1450 - 1800$ & $2.991 - 3.652$ & $2.079 - 2.595$ & $1.302 - 1.674$ \\[-1ex] 
                            & & $1940 - 2450$ & -- & -- & $2.079 - 2.640$ \\
   \hline
  \end{tabular} 
  \begin{tablenotes}
   \item $\mathrm{^a}$This band is divided in two by the gap in atmospheric transmission between $J$ band and $H$ band.   
   \item $\mathrm{^b}$This band is divided in two by the gap in atmospheric transmission between $H$ band and $K$ band. 
   \item $\mathrm{^c}$The red end of this band is truncated from 1369 nm to 1350 nm due to the gap in atmospheric transmission between $J$ band and $H$ band.
  \end{tablenotes}
 \end{threeparttable}
\end{table*}

\section{IFS instruments and accessible emission line suites}
\label{sec:ins_z}

In this section, we briefly introduce the three future IFS instruments on which we focus in this work in \autoref{sec:ins}, and then in \autoref{sec:z_range} define emission-line suites that are accessible using single bands of these instruments at a range of redshift windows.

\subsection{IFS instruments}
\label{sec:ins}

\subsubsection{MAVIS}

MAVIS \citep{rigaut2020,rigaut2021} is a general-purpose instrument that will operate on the AO facility of the Very Large Telescope (VLT) and includes three modules: an AO module, an imager module, and an IFS module. The technical field of MAIVS covers 120$''$ in diameter, from which up to 3 natural guide stars are selected and fed to AO module to correct the wavefront. The IFS module comprises a fine spatial mode with a $2\farcs5\times3\farcs6$ FoV and sampled by 25 mas pixels, and a coarse spatial mode with a $5\farcs0\times7\farcs2$ FoV sampled by 50 mas pixels. The spectral configuration comprises blue and red bands. The low resolution modes have a median $R\approx 5,900$ with a continuous coverage from 370 nm to 1000 nm split across two separate bands (LR-Red and LR-Blue). The high resolution mode has a median $R\approx 14,700$ in the blue band and $R\approx 11,500$ in the red band, with coverage of 425-550 nm and 630-880 nm, respectively. In this work we will focus on the two low-resolution spectral modes, as they provide the broadest simultaneous wavelength coverage. We list these bands and their wavelength ranges in the left three columns of \autoref{tab:redshift}.

\subsubsection{GMTIFS}

GMTIFS \citep{mcgregor2012} is one of six instruments planned for the Giant Magellan Telescope (GMT) and contains an IFS, an imager, and an on-instrument wave-front sensor that is used to correct the wavefront with the laser tomography AO system. The correction is set to milli-arcsecond precision corresponding to $\sim 1~\mu$m in the focal plane. The IFS comprises four FoVs from $4\farcs 40\times 2\farcs 25$ to $0\farcs 528\times0\farcs 27$ with square spaxel sizes from 50 mas to 6 mas. The spectrograph consists of a medium resolution mode of $R\approx 5,000$ and a high resolution mode of $R\approx 10,000$. The medium resolution mode comprises three bands: $mZJ$ ($0.89-1.35~\mu m$), $mJH$ ($1.19-1.80~\mu m$), and $mHK$ ($1.64-2.49~\mu m$). The high resolution mode comprises four bands: $hZ$ ($0.92-1.13~\mu m$), $hJ$ ($1.10-1.35~\mu m$), $hH$ ($1.47-1.80~\mu m$), and $hK$ ($2.04-2.51~\mu m$). As with MAVIS, in this work we will focus on the three medium-resolution modes due to their broad wavelength ranges; we list these in \autoref{tab:redshift}.

\subsubsection{HARMONI}

HARMONI \citep{thatte2021} is the visible and near-infrared (NIR) AO-assisted IFS for the Extremely Large Telescope (ELT); it will achieve near diffraction-limited spatial resolution in the NIR, but in the visible bands is seeing-limited. The spaxel size for visible or non-AO observation is 30 mas $\times$ 60 mas with a FoV of $9\farcs 12 \times 6\farcs 12$. For the NIR bands with AO wavefront correction, three square spaxel sizes from 20 mas to 6 mas are provided with FoV sizes from $3\farcs 04 \times 4\farcs 08$ to $0\farcs 61 \times 0\farcs 82$ to allow the user to optimise the balance between resolution and sensitivity. HARMONI comprises eleven gratings covering the wavelength ranges from visible to $K$ band with spectral resolution settings of $R\approx$ 3,000, 7,000, and 17,000. The low resolution gratings comprise three bands: $V+R$ ($0.458-0.820~\mu m$), $Iz+J$ ($0.811-1.369~\mu m$), and $H+K$ ($1.450-2.450~\mu m$). The medium resolution gratings comprise four bands: $Iz$ ($0.830-1.050~\mu m$), $J$ ($1.046-1.324~\mu m$), $H$ ($1.435-1.815~\mu m$), and $K$ ($1.951-2.469~\mu m$). The high resolution gratings comprise four bands: $z$-high ($0.828-0.902~\mu m$), $H$-high ($1.538-1.678~\mu m$), $K$-short ($2.017-2.201~\mu m$), and $K$-long ($2.199-2.400~\mu m$). We again focus on the three low-resolution modes in this work, and list their properties in \autoref{tab:redshift}.

\subsection{Line suites and redshift ranges}
\label{sec:z_range}

\begin{figure}
    {\includegraphics[width=\columnwidth]{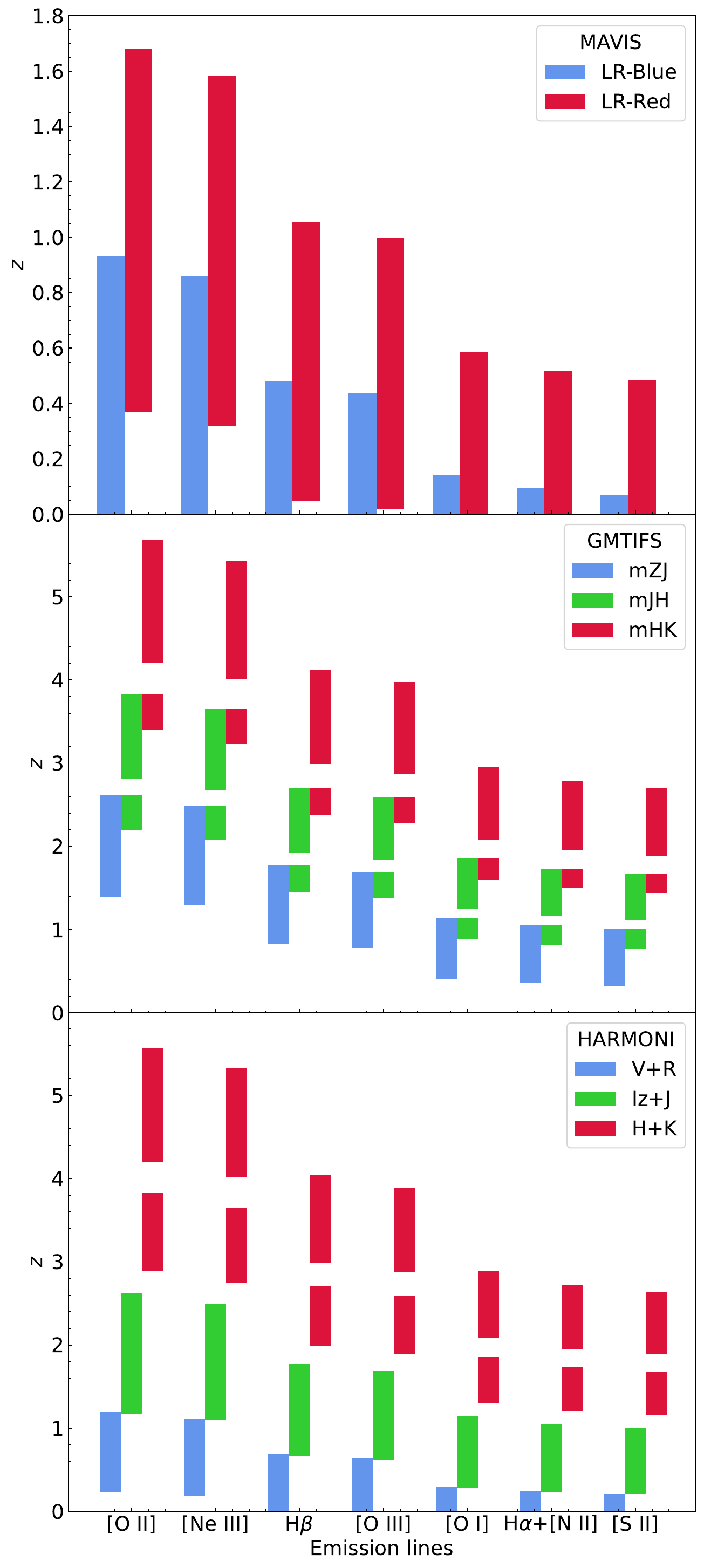}}
    \caption{The redshift ranges over which each of the emission lines that we use in this work is accessible for each observing band of MAVIS (top panel), GMTIFS (middle panel), and HARMONI (bottom panel). The wavelength coverages of the bands shown are listed in \autoref{tab:redshift}. For the purpose of reducing clutter, we have combined the [O~\textsc{ii}]$~\lambda\lambda3727, 29$ and [S~\textsc{ii}]$~\lambda\lambda6717, 31$ doublets, as well as the very closely-spaced line pair H$\alpha$ and [N~\textsc{ii}]$~\lambda6584$, into single lines denoted as [O~\textsc{ii}], [S~\textsc{ii}], and H$\alpha$+[N~\textsc{ii}] in the legend, respectively. The redshift ranges shown for these line pairs are those over which both lines can be observed.}
    \label{fig:line_redshift}
\end{figure}

\begin{figure}
    {\includegraphics[width=\columnwidth]{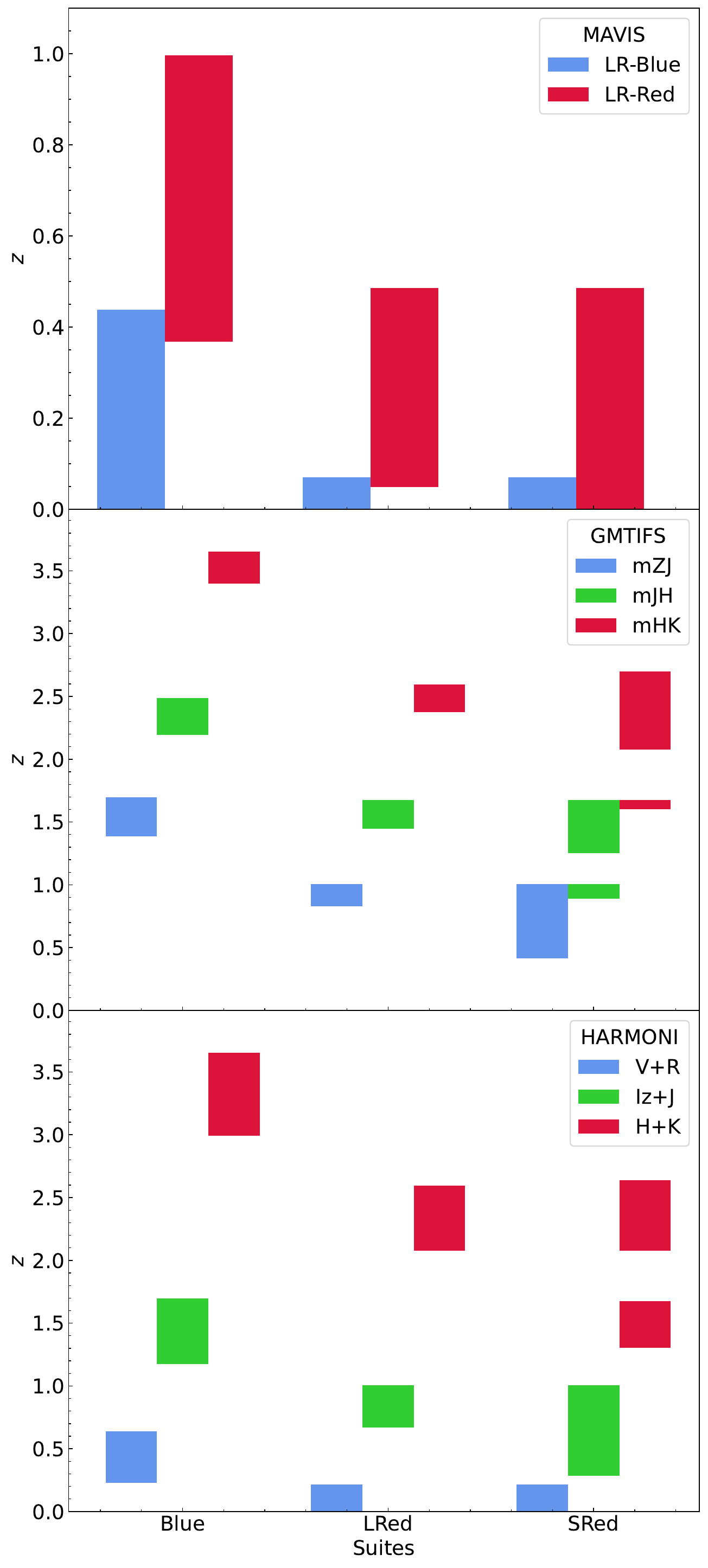}}
    \caption{Same as \autoref{fig:line_redshift}, but now showing the accessible redshift ranges for line suites rather than individual lines. We also list these redshift ranges in \autoref{tab:redshift}.}
    \label{fig:suites_redshift}
\end{figure}

As described above, the three instruments considered here include a range of different spectral modes that can only be used one at a time, and thus only a subset of the optical emission lines that we use as metallicity diagnostics at $z=0$ will be available in a single exposure. Which subsets are accessible depends on both the observing band chosen and the redshift of the target. The first step of this work is to understand, for each specific subset of emission lines that we might consider using to measure metallicity, over what redshift range all the required lines in each subset can be captured by a single-band observation.

In this work we consider ten rest-band optical emission lines that are frequently used in metallicity diagnostics for both AGN-host and non-host galaxies: the [O~\textsc{ii}]$~\lambda\lambda3727,29$ doublet, [Ne~\textsc{iii}]$~\lambda3869$, H$\beta$, [O~\textsc{iii}]$~\lambda5007$, [O~\textsc{i}]$~\lambda6300$, H$\alpha$, [N~\textsc{ii}]$~\lambda6584$, and the [S~\textsc{ii}]$~\lambda\lambda6717,31$ doublet\footnote{Note that if we were also to include the [S~\textsc{iii}] $\lambda\lambda 9068, 9530$ doublet, we could in principle constrain the ionization parameter by combining this with [S~\textsc{ii}] doublet at the red end of the optical band. However, we do not consider this option for two reasons, one practical and one based on the potential utility of this line. The practical reason is that this doublet is not included in the the MPA-JHU catalog from which we select our sample in this work. The reason based on potential utility is that this doublet is far away from all the other lines, and as a result we find that diagnostics that rely on this doublet to constrain the ionization parameter but that also need other, bluer lines to diagnose the metallicity, would be usable only over an extremely narrow range of redshifts for our three target instruments. This would greatly limit the potential utility of diagnostics that require [S~\textsc{iii}] $\lambda\lambda 9068, 9530$, and for this reason we will not consider them further.}. This list includes all the traditional optical strong lines for star-forming galaxies, together with the high excitation lines that are sensitive to AGN emission. \autoref{fig:line_redshift} shows the redshift range over which each of these lines falls within the available bands of MAVIS (top panel), GMTIFS (middle panel), and HARMONI (bottom panel). There are gaps in the $mJH$ and $mHK$ bands of GMTIFS, and $H+K$ band of HARMONI due to gaps in  atmospheric transmission  between $J$, $H$, and $K$ bands; for the purposes of our tables and figures, we place the edges of these gaps at the wavelengths where atmospheric transmission falls below 50\%. 

From \autoref{fig:line_redshift} it is obvious that these emission lines cluster into three groups due to their relative wavelength separations: 1) the [O~\textsc{ii}] doublet and [Ne~\textsc{iii}]; 2) H$\beta$ and [O~\textsc{iii}]; and 3) [O~\textsc{i}], H$\alpha$, [N~\textsc{ii}], and the [S~\textsc{ii}] doublet. Based on this clustering, it is natural to group the lines into three suites, which will frequently be available together:
\begin{itemize}
    \item \textbf{Blue}: This suite includes the [O~\textsc{ii}] doublet, [Ne~\textsc{iii}], H$\beta$ and [O~\textsc{iii}]. 

    \item \textbf{LRed} (Long Red): This suite includes H$\beta$, [O~\textsc{iii}], [O~\textsc{i}], H$\alpha$, [N~\textsc{ii}], and the [S~\textsc{ii}] doublet.

    \item \textbf{SRed} (Short Red): This suite includes [O~\textsc{i}], H$\alpha$, [N~\textsc{ii}], and the [S~\textsc{ii}] doublet.
\end{itemize}
We choose these suites not just because they are close together in wavelength, but in order to ensure that each contains at least one line ratio that is known to be sensitive to metallicity \citep[according to][]{kewley2019b}; for example, the Blue suite contains [O~\textsc{ii}], [O~\textsc{iii}], and H$\beta$, the lines used to form the classical $R_{23}$ calibrator \citep{pagel1979}. The combination of the Blue and LRed suites and the combination of the Blue and SRed suites both correspond to all of the optical lines typically used at $z=0$.

\autoref{fig:suites_redshift} shows the redshift ranges over which each emission-line \textit{suite} is accessible for each instrument and band; we also list these in the three right columns of \autoref{tab:redshift}. Notice that, as expected, there are only very limited redshift ranges where both the Blue and LRed suites (i.e., the full optical suite) are accessible with a single-band observation: 0-0.07 and 0.368-0.486 in the LR-Blue and LR-Red bands of MAVIS, and none for GMTIFS and HARMONI. Outside these windows, and thus at all $z > 0.486$, capturing all the optical lines of interest would either require observations in multiple bands (e.g., using GMTIFS $mZJ$ and $mJH$ to capture the Blue and LRed suites, respectively, at $z\sim 1.5$), or is entirely impossible given the available instruments (e.g., at $z\sim 2.1$ only the SRed band is observable by GMTIFS and only the LRed or SRed from HARMONI, with Blue inaccessible to either instrument). This further motivates our approach of investigating the efficacy of using single-band observations to obtain metallicities.

\section{Data and analysis methods}
\label{sec:met_cali}

Here we introduce the sample of local galaxies that we will use for our experiment in \autoref{sec:sample}, and then give a brief introduction to the \nebbayes~code that we will use to derive metallicities for these galaxies from various line suites in \autoref{sec:nb}.

\subsection{Sample selection}
\label{sec:sample}

Our goal is to understand how using only a subset of optical emission lines, rather than the full line suite to which we have access at $z=0$, impacts metallicity measurements in both non-AGN and AGN-host galaxies. We must therefore define a sample of galaxies on which to carry out these tests. For this purpose we select galaxies from SDSS data release (DR) 7 \citep{abazajian2009}. The emission line fluxes for galaxies in this sample are measured from SDSS single fiber spectra covering the central 3$''$ of each galaxy \citep{tremonti2004}, and extracted as part of the MPA-JHU spectral analysis data release available at \url{https://wwwmpa.mpa-garching.mpg.de/SDSS/DR7/}. The Bayesian method we use to convert line fluxes to metallicities (see \autoref{sec:nb}) requires measurement uncertainties as well as central values. We therefore scale the flux errors by empirical factors recommended as part of the data release -- see \url{https://wwwmpa.mpa-garching.mpg.de/SDSS/DR7/raw_data.html}.

From this parent data set, we obtain a sample using selection criteria similar to those of \citet{thomas2019}. Specifically, we require that 
\begin{itemize}
\item All 10 emission lines used in this work (see \autoref{sec:z_range}) have positive fluxes and flux errors.
\item The signal-to-noise ratio (SNR) of the H$\beta$ line is greater than 5. We select on the SNR of the Balmer line only, instead of separately on each of metal lines we use, in order to avoid biasing against low metallicity systems \citep{salim2014,baker2023}.
\item The equivalent width (EW) of H$\alpha$ is greater than 3 \AA. This requirement is to remove the contamination from post-asymptotic giant branch (pAGB) stars \citep{cid2011,belfiore2016,lacerda2018}.
\item The H$\alpha/$H$\beta$ flux ratio is greater than 2.7. This condition excludes galaxies with unphysical Balmer decrements.
\item We exclude systems for which the spectral type classification \texttt{spectrotype} is `QSO' or `AGN BROADLINE'; this selection is to reduce the flux contamination from broad-line regions.
\end{itemize}

Applying these conditions yields a sample of approximately 200,000 galaxies. In order to produce separate AGN and non-AGN subsamples, we classify each of these systems based on their position in the BPT diagram \citep{baldwin1981}. We classify galaxies located below the \citet{kauffmann2003} empirical demarcation line as BPT-SF (non-AGN), and galaxies located above \citet{kewley2001} theoretical maximum starburst line as AGN-host. Applying these classifications yields a sample of 187,955 non-AGN (BPT-SF) galaxies and 9,086 AGN-host galaxies. Because we must Monte-Carlo-resample each galaxy multiple times in order to account for uncertainties, applying our full analysis pipeline a sample of this size would be quite computationally expensive. We therefore randomly select 10,000 BPT-SF galaxies from the total 187,955 available, yielding a sample of similar size to the AGN-host galaxies, that we will use for the remainder of this work.

\subsection{Metallicity estimation with \nebbayes}
\label{sec:nb}

For the purposes of determining the metallicities of our sample galaxies we use \nebbayes~ \citep{thomas2018}, a Bayesian code for estimating a series of posterior parameters -- including metallicity -- by comparing the emission-line fluxes with theoretical model grids generated by photoionization models such as \mappings~\citep{sutherland2017} or \textsc{cloudy}~\citep{Chatzikos2023}. We use \nebbayes~for our experiment because it offers two key advantages: first, as discussed in \autoref{sec:intro}, \nebbayes~can cope with spectra that contain a mixture of AGN-driven and stellar-driven ionised regions, allowing us to achieve our goal of comparing AGN and non-AGN host galaxies on equal footing. Second, because it is a Bayesian method, \nebbayes~automatically and naturally copes with missing data, allowing us to carry out the experiment of seeing how inferred metallicities change as we gradually remove line data in a single, self-consistent framework. Here we only briefly introduce \nebbayes~and the parameter choices we make when running it, and refer the readers to \citet{li2024} for details.

\nebbayes~takes as its basic input two grids of nebular emission spectra, one computed for $\hii$~regions driven by stellar-like ionizing spectra and another computed for narrow-line regions (NLRs) driven by AGN-like ionizing spectra. We use \mappings~to generate both these sets of grids. For the $\hii$ region case, the grid has three dimensions: gas pressure $P$, ionization parameter $U_{\hii}$, and oxygen abundance $\mathrm{O}/\mathrm{H}$. We sample these uniformly in log space, with 20 points of gas pressure $\log (P/k) \in [4,8]$, 40 points of ionization parameter $\log U_{\hii} \in [-4, -2]$, and 160 points of metallicity $\Oabu \in [7.75, 9.252]$. We do not assume simple Solar-scaled abundances for elements other than oxygen in the nebula; instead, we set the abundances relative to oxygen using the element-by-element ``Galactic Concordance'' scaling provided by \citet{nicholls2017}, which is calibrated based on observations of abundance patterns in nearby nebulae and stars. In particularly, the nitrogen-to-oxygen abundance ratio as a function of $\Oabu$ is fitted with a combination of two linear functions that address both primary and secondary nitrogen. These non-Solar-scaled abundances are particularly important for the [N~\textsc{ii}]$/$[O~\textsc{ii}] line ratio. However, we caution that even though our use of ``Galactic Concordance'' abundances allows us to capture variation of mean N/O ratio with metallicity, it does not fully capture the observed scatter whereby, at fixed metallicity, the N/O can vary by as much as a factor of 10 from galaxy to galaxy. This spread introduces unavoidable systematic uncertainties into any metallicity analysis, and is one of several that we attempt to capture by, following the standard procedure for running \nebbayes~\citep{thomas2018}, adding a uniform fractional systematic error of 35\% to its predictions. Finally, to compute the ionizing spectrum fed to \mappings, we use \textsc{fsps} \citep[Flexible Stellar Population Synthesis,][]{conroy2009,conroy2010}; for this calculation we adopt the BaSeL stellar library \citep{westera2003} and Stromlo Stellar Tracks \citep{grasha2021}, which are based on the same Galactic Concordance abundances we assume for nebular gas. For galaxies that we classify as BPT-SF, meaning that there is no evidence for an AGN contribution, we save computational time by using these grids alone when estimating metallicity.

For galaxies that do show evidence for an AGN contribution, we use grids that combine $\hii$ region and NLR contributions, and we therefore require grids of photoionization models for the latter. We use \textsc{oxaf} \citep{thomas2016} to generate the theoretical ionizing spectra that are required as input to these calculations. These spectra are described by three parameters: the power-law index $\Gamma$ of the non-thermal component, the fraction of this non-thermal component to the total AGN luminosity $p_\mathrm{NT}$, and the energy peak of the Big Blue Bump component from the AGN accretion disc $\log (E_\mathrm{peak}/\mathrm{keV})$. Following \citet{thomas2019}, who found that varying these parameters had minimal effects on the inferred metallicity, we fix $\Gamma=-2.0$, $p_\mathrm{NT}=0.15$, and $\log (E_\mathrm{peak}/\mathrm{keV}) = 1.35$. Thus the NLR grids also have three dimensions: $\log (P/k)$, ionization parameter $\log U_\mathrm{NLR}$, and $\Oabu$. We sample pressure and oxygen abundance on the same grid that we use for the $\hii$-region grid, while for ionization parameter we use 40 sample points uniformly distributed in the range $\log U_\mathrm{NLR} \in [-3.8, -0.2]$. We then generate the combined grids including both the $\hii$ region and NLR contributions by introducing $f_{\hii}$, which means the fraction of the H$\alpha$ flux coming from $\hii$ regions. When combining two grids, we match the gas pressure $\log (P/k)$ and the metallicity $\Oabu$, and fix the $\hii$ region ionization parameter to $\log U_{\hii}=-3.25$ \citep[again following][]{thomas2019}. The combined grids thus comprise four free parameters: $\log (P/k)$, $\log U_\mathrm{NLR}$, $\Oabu$, and $f_{\hii} \in [0, 1]$; we sample this last dimension with 40 points uniformly in $f_{\hii}$.

The fiducial metallicity that we will use as the ``ground truth'' against which to compare inference from our reduced line suites is estimated from all ten of the emission lines listed in \autoref{sec:z_range}. In addition to these inputs, \nebbayes~ offers an `emission-line prior' feature that uses emission-line ratios that are sensitive to one specific free parameter to constrain the posterior distributions; in effect, this one line ratio is over-weighted in the posteriors computed for that quantity -- see \citet{thomas2018} for details. In our fiducial case, we use the line ratios [S~\textsc{ii}]$~\lambda 6716/$[S~\textsc{ii}]$~\lambda 6731$ for our priors on gas pressure \citep{kewley2019a}, and [N~\textsc{ii}]$/$[O~\textsc{ii}] for our priors on  metallcity \citep{kewley2002}, following the procedure used in \citet{li2024} and recommended by \citet{thomas2019}. In practice, what this means is that we generate theoretical [S~\textsc{ii}]$~\lambda 6716/$[S~\textsc{ii}]$~\lambda 6731$ and [N~\textsc{ii}]$/$[O~\textsc{ii}] ratios at each model grid point along with line fluxes. We also calculate these two line ratios of observational spectra, propagating the flux uncertainties to line ratio uncertainties. We finally input these line ratios in addition to fluxes into \nebbayes~to estimate the posterior distributions. In \autoref{sec:analysis}, we apply this procedure to all available line ratios in each line suite sensitive to either metallicity or gas pressure to select the one with the best performance. We will adopt a naming convention for combinations of line suite and prior of the form \texttt{LineSuite\_Prior1\_Prior2} to indicate each run, and thus
we assign as the name of this fiducial combination \texttt{All\_NO\_S2}. We discuss the set of priors that we consider for other line suites below.

Used on either BPT-SF galaxies with the $\hii$ region-only grid or on AGN-host galaxies with the combined grid, the output of \nebbayes~is a set of posterior probabilities evaluated at each point in our model grid. We take the mode of these posteriors as the metallicity estimate for each galaxy. To marginalise over observational uncertainties, in all cases we repeat this procedure 20 times, each time using a different realization of the flux of each line drawn from a Gaussian distribution centered on the best-estimated flux for that line and with a dispersion equal to the uncertainty, thereby generating 20 different metallicity estimates per galaxy.\footnote{Our sampling procedure raises the complication of how to handle realizations where a given flux measurement is negative, since \nebbayes~requires positive fluxes \citep{thomas2019}. For this purpose we discard and redraw realizations where any of the strong-line fluxes (the [O~\textsc{ii}] doublet, H$\beta$, [O~\textsc{iii}], H$\alpha$, [N~\textsc{ii}], and the [S~\textsc{ii}] doublet) are negative, and for cases where one of the weaker lines ([Ne~\textsc{iii}] and [O~\textsc{i}]) has a negative flux, we set the flux to a positive value equal to 5 per cent of the corresponding flux error. These choices have little impact on the results, since they only affect the minority of galaxies where these lines are only marginally detected, and since the weak lines have low weight on the final posterior distribution.}

There is one final complication that we encounter when applying \nebbayes~to our reduced line suites: dereddening. \nebbayes~ handles this by taking the Balmer decrement H$\alpha/$H$\beta$ of each grid point as the intrinsic ratio, and dereddening other lines using the observed H$\alpha/$H$\beta$ and the dust attenuation curve adopted from \citet{vogt2013}. The H$\alpha/$H$\beta$ of the best-fitting grid point is then taken to be the real intrinsic ratio. Our complication is that, of our three candidate line suites, only LRed contains both H$\alpha$ and H$\beta$, which means that observations using other line suites will need to adopt a different strategy for de-reddening. Since there are many possible approaches to this problem, and our goal is to survey metallicity measurement methods and not dust correction ones, we ignore this complication by assuming that all galaxies can be adequately dereddened by some means. To this end, we only let \nebbayes~ do dereddening on the fiducial run \texttt{All\_NO\_S2}, and for all other line suites and prior combinations, we use the inferred intrinsic H$\alpha/$H$\beta$ ratio from the fiducial case to ``pre-deredden'' the emission lines by making the reddening correction on the data before feeding them to \nebbayes, and then simply disabling \nebbayes's reddening correction capability. In addition to simplicity, this procedure has the advantage that it isolates the problem of metallicity inference from that of reddening correction, since our procedure forces all cases to be computed using the same reddening correction.

\section{Performance of different line suites}
\label{sec:analysis}

\begin{table*} 
 \centering
 \renewcommand{\arraystretch}{1.3}
 \begin{threeparttable}
  \caption{The combinations of emission-line suite and line ratio prior weighting used to constrain the posterior distributions in this work.}
  \vspace{4mm}
  \label{tab:comb_names}

  \begin{tabular}{p{2cm}p{7.4cm}p{6cm}}
   \hline
   Name & \multicolumn{1}{>{\centering\arraybackslash}m{8cm}}{Emission lines} & \multicolumn{1}{>{\centering\arraybackslash}m{6cm}}{Line ratios weights}\\
   \hline
   \texttt{All\_NO\_S2} & [O~\textsc{ii}] doublet, [Ne~\textsc{iii}], H$\beta$, [O~\textsc{iii}], [O~\textsc{i}], H$\alpha$, [N~\textsc{ii}], [S~\textsc{ii}] doublet & [N~\textsc{ii}]$/$[O~\textsc{ii}], [S~\textsc{ii}]$~\lambda 6716/$[S~\textsc{ii}]$~\lambda 6731$ \\
   \hline
   \texttt{Blue} & [O~\textsc{ii}] doublet, [Ne~\textsc{iii}], H$\beta$, [O~\textsc{iii}] & -- \\
   \texttt{Blue\_O2} & [O~\textsc{ii}] doublet, [Ne~\textsc{iii}], H$\beta$, [O~\textsc{iii}] & [O~\textsc{ii}]$~\lambda 3729/$[O~\textsc{ii}]$~\lambda 3726$ \\
   \texttt{Blue\_O3\_O2} & [O~\textsc{ii}] doublet, [Ne~\textsc{iii}], H$\beta$, [O~\textsc{iii}] & [O~\textsc{iii}]$/$H$\beta$, [O~\textsc{ii}]$~\lambda 3729/$[O~\textsc{ii}]$~\lambda 3726$ \\
   \texttt{Blue\_O2\_O2} & [O~\textsc{ii}] doublet, [Ne~\textsc{iii}], H$\beta$, [O~\textsc{iii}] & [O~\textsc{ii}]$/$H$\beta$, [O~\textsc{ii}]$~\lambda 3729/$[O~\textsc{ii}]$~\lambda 3726$ \\
   \texttt{Blue\_R23\_O2} & [O~\textsc{ii}] doublet, [Ne~\textsc{iii}], H$\beta$, [O~\textsc{iii}] & ([O~\textsc{ii}]+[O~\textsc{iii}])$/$H$\beta$, [O~\textsc{ii}]$~\lambda 3729/$[O~\textsc{ii}]$~\lambda 3726$ \\
   \hline
   \texttt{LRed} & H$\beta$, [O~\textsc{iii}], [O~\textsc{i}], H$\alpha$, [N~\textsc{ii}], [S~\textsc{ii}] doublet & -- \\
   \texttt{LRed\_S2} & H$\beta$, [O~\textsc{iii}], [O~\textsc{i}], H$\alpha$, [N~\textsc{ii}], [S~\textsc{ii}] doublet & [S~\textsc{ii}]$~\lambda 6716/$[S~\textsc{ii}]$~\lambda 6731$ \\
   \texttt{LRed\_S2\_S2} & H$\beta$, [O~\textsc{iii}], [O~\textsc{i}], H$\alpha$, [N~\textsc{ii}], [S~\textsc{ii}] doublet & [S~\textsc{ii}]$/$H$\alpha$, [S~\textsc{ii}]$~\lambda 6716/$[S~\textsc{ii}]$~\lambda 6731$ \\
   \texttt{LRed\_N2\_S2} & H$\beta$, [O~\textsc{iii}], [O~\textsc{i}], H$\alpha$, [N~\textsc{ii}], [S~\textsc{ii}] doublet & [N~\textsc{ii}]$/$H$\alpha$, [S~\textsc{ii}]$~\lambda 6716/$[S~\textsc{ii}]$~\lambda 6731$ \\
   \texttt{LRed\_NS\_S2} & H$\beta$, [O~\textsc{iii}], [O~\textsc{i}], H$\alpha$, [N~\textsc{ii}], [S~\textsc{ii}] doublet & [N~\textsc{ii}]$/$[S~\textsc{ii}], [S~\textsc{ii}]$~\lambda 6716/$[S~\textsc{ii}]$~\lambda 6731$ \\
   \texttt{LRed\_O3\_S2} & H$\beta$, [O~\textsc{iii}], [O~\textsc{i}], H$\alpha$, [N~\textsc{ii}], [S~\textsc{ii}] doublet & [O~\textsc{iii}]$/$H$\beta$, [S~\textsc{ii}]$~\lambda 6716/$[S~\textsc{ii}]$~\lambda 6731$ \\
   \texttt{LRed\_O3N2\_S2} & H$\beta$, [O~\textsc{iii}], [O~\textsc{i}], H$\alpha$, [N~\textsc{ii}], [S~\textsc{ii}] doublet & ([O~\textsc{iii}]$/$H$\beta$)$/$([N~\textsc{ii}]$/$H$\alpha$), [S~\textsc{ii}]$~\lambda 6716/$[S~\textsc{ii}]$~\lambda 6731$ \\
   \hline 
   \texttt{SRed} & [O~\textsc{i}], H$\alpha$, [N~\textsc{ii}], [S~\textsc{ii}] doublet & -- \\
   \texttt{SRed\_S2} & [O~\textsc{i}], H$\alpha$, [N~\textsc{ii}], [S~\textsc{ii}] doublet & [S~\textsc{ii}]$~\lambda 6716/$[S~\textsc{ii}]$~\lambda 6731$ \\
   \texttt{SRed\_S2\_S2} & [O~\textsc{i}], H$\alpha$, [N~\textsc{ii}], [S~\textsc{ii}] doublet & [S~\textsc{ii}]$/$H$\alpha$, [S~\textsc{ii}]$~\lambda 6716/$[S~\textsc{ii}]$~\lambda 6731$ \\
   \texttt{SRed\_N2\_S2} & [O~\textsc{i}], H$\alpha$, [N~\textsc{ii}], [S~\textsc{ii}] doublet & [N~\textsc{ii}]$/$H$\alpha$, [S~\textsc{ii}]$~\lambda 6716/$[S~\textsc{ii}]$~\lambda 6731$ \\
   \texttt{SRed\_NS\_S2} & [O~\textsc{i}], H$\alpha$, [N~\textsc{ii}], [S~\textsc{ii}] doublet &[N~\textsc{ii}]$/$[S~\textsc{ii}], [S~\textsc{ii}]$~\lambda 6716/$[S~\textsc{ii}]$~\lambda 6731$ \\
   \hline       
  \end{tabular} 
 \end{threeparttable}
\end{table*}

\begin{table*} 
 \centering
 \renewcommand{\arraystretch}{1.3}
 \begin{threeparttable}
  \caption{Performance statistics for each emission line suite and weighting scheme listed in \autoref{tab:comb_names}. The combination name with an asterisk marks the best-performing weighting scheme for the corresponding emission line suite for BPT-SF galaxies.}
  \vspace{1mm}
  \label{tab:statistics_sf}

  \begin{tabular}{lcccc}
   \hline
   Name & Pearson coefficient & Median orthogonal distance (dex) & RMS orthogonal distance (dex) & Valid fraction \\
   \hline
   \texttt{Blue}$^*$ & 0.892 & 0.060 & 0.073 & 99.3\% \\
   \texttt{Blue\_O2} & 0.875 & 0.073 & 0.073 & 98.1\% \\
   \texttt{Blue\_O3\_O2} & 0.879 & 0.073 & 0.073 & 98.3\% \\
   \texttt{Blue\_O2\_O2} & 0.870 & 0.073 & 0.074 & 98.0\% \\
   \texttt{Blue\_R23\_O2} & 0.873 & 0.073 & 0.074 & 98.2\% \\
   \hline
   \texttt{LRed} & 0.747 & -0.013 & 0.117 & 99.4\% \\
   \texttt{LRed\_S2} & 0.752 & -0.013 & 0.116 & 99.3\% \\
   \texttt{LRed\_S2\_S2} & 0.710 & -0.007 & 0.124 & 99.1\% \\
   \texttt{LRed\_N2\_S2} & 0.743 & -0.060 & 0.131 & 99.1\% \\
   \texttt{LRed\_NS\_S2} & 0.784 & -0.053 & 0.119 & 99.3\% \\
   \texttt{LRed\_O3\_S2} & 0.753 & 0.000 & 0.118 & 99.4\% \\
   \texttt{LRed\_O3N2\_S2}$^*$ & 0.826 & -0.020 & 0.102 & 99.3\% \\
   \hline
   \texttt{SRed} & 0.792 & -0.134 & 0.105 & 98.2\% \\
   \texttt{SRed\_S2} & 0.814 & -0.140 & 0.095 & 99.3\% \\
   \texttt{SRed\_S2\_S2} & 0.315 & 0.040 & 0.161 & 99.9\% \\
   \texttt{SRed\_N2\_S2} & 0.837 & 0.060 & 0.084 & 98.1\% \\
   \texttt{SRed\_NS\_S2}$^*$ & 0.844 & -0.007 & 0.087 & 99.8\% \\
   \hline   
  \end{tabular} 
 \end{threeparttable}
\end{table*}

\begin{table*} 
 \centering
 \renewcommand{\arraystretch}{1.3}
 \begin{threeparttable}
  \caption{The same as \autoref{tab:statistics_sf}, but for AGN-host galaxies.}
  \vspace{1mm}
  \label{tab:statistics_agn}

  \begin{tabular}{lcccc}
   \hline
   Name & Pearson coefficient & Median orthogonal distance (dex) & RMS orthogonal distance (dex) & Valid fraction \\
   \hline
   \texttt{Blue}$^*$ & 0.536 & -0.060 & 0.124 & 80.0\% \\
   \texttt{Blue\_O2} & 0.533 & -0.053 & 0.127 & 80.0\% \\
   \texttt{Blue\_O3\_O2} & 0.525 & -0.047 & 0.125 & 79.9\% \\
   \texttt{Blue\_O2\_O2} & 0.531 & -0.060 & 0.125 & 86.1\% \\
   \texttt{Blue\_R23\_O2} & 0.524 & -0.047 & 0.125 & 81.7\% \\
   \hline
   \texttt{LRed} & 0.504 & -0.033 & 0.102 & 88.8\% \\
   \texttt{LRed\_S2} & 0.500 & -0.033 & 0.103 & 87.9\% \\
   \texttt{LRed\_S2\_S2} & 0.488 & -0.033 & 0.105 & 87.8\% \\
   \texttt{LRed\_N2\_S2} & 0.526 & -0.040 & 0.099 & 87.9\% \\
   \texttt{LRed\_NS\_S2}$^*$ & 0.556 & -0.040 & 0.099 & 88.8\% \\
   \texttt{LRed\_O3\_S2} & 0.503 & -0.047 & 0.101 & 89.1\% \\
   \texttt{LRed\_O3N2\_S2} & 0.517 & -0.040 & 0.098 & 86.9\% \\
   \hline
   \texttt{SRed}$^*$ & 0.514 & 0.007 & 0.105 & 94.4\% \\
   \texttt{SRed\_S2} & 0.478 & -0.027 & 0.109 & 93.0\% \\
   \texttt{SRed\_S2\_S2} & 0.375 & -0.013 & 0.088 & 99.4\% \\
   \texttt{SRed\_N2\_S2} & 0.568 & 0.147 & 0.081 & 35.9\% \\
   \texttt{SRed\_NS\_S2} & 0.524 & 0.160 & 0.094 & 20.4\% \\
   \hline   
  \end{tabular} 
 \end{threeparttable}
\end{table*}

\begin{figure*}
    \resizebox{17cm}{!}{\includegraphics{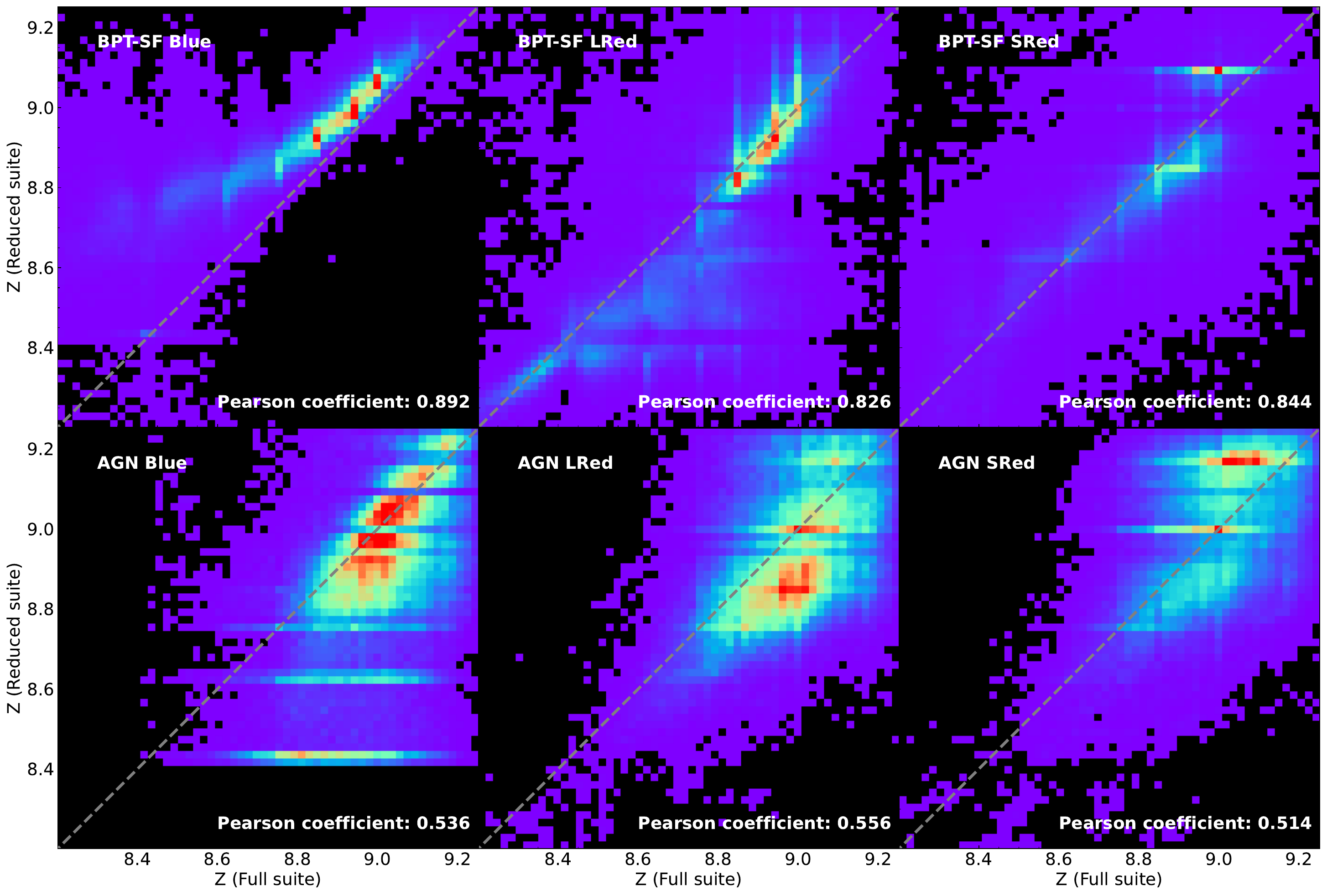}}
    \caption{Comparisons between metallicities derived from the best-performing line ratio weighting scheme for each emission line suite (as indicated by the asterisks in \autoref{tab:statistics_sf} and \autoref{tab:statistics_agn}) and the fiducial metallicity estimated from all lines. From left to right, columns show the Blue, LRed, and SRed suites. The top row shows BPT-SF galaxies, and bottom row shows AGN-host galaxies. Each panel shows a 2D histogram of the best-fit metallicity of each galaxy, with metallicities derived from our fiducial estimator \texttt{All\_NO\_S2} on the horizontal axis and from the indicated line suite on the vertical axis. The counts of each bin are normalised to the maximum of each panel. From purple to red, the histograms show the range from 0 to 0.9, with black indicating a value of exactly zero (i.e., no samples fall into that bin). The Pearson correlation coefficients of each comparison are listed at the bottom-right corner of each panel. We refer readers to \autoref{tab:statistics_sf} and \autoref{tab:statistics_agn} for more statistics.}
    \label{fig:performance}
\end{figure*}

We are now prepared to examine the performance of three line suites introduced in \autoref{sec:z_range} by comparing metallicities derived with them to the fiducial metallicity derived using all emission lines, i.e., via the \texttt{All\_NO\_S2} combination introduced in \autoref{sec:nb}. For each line suite we consider several possible line ratio prior weighting strategies (see \autoref{sec:nb} for details). The full set of line suites and weighting combinations we test is listed in \autoref{tab:comb_names}.

In order to evaluate the various options both qualitatively and quantitatively we use two tools. The first is ``metallicity-metallicity'' plots, which place fiducial metallicity on the horizontal axis and the metallicity estimated using one of the candidate line suites and weightings on the vertical axis, and then shows a 2d heatmap of the distribution of galaxies in this plane. We generate these heatmaps from our 20 Monte Carlo realisations per sample galaxy to ensure that we are properly sampling the uncertainty distribution. Ideal performance on this plot consists of all galaxies clustering along the 1-to-1 line. We provide metallicity-metallicity plots for all the possible line suites and weighting combinations that we discuss below in \aref{app:performance_all}, and in the main text we will focus only on the best-performing options.

Our second method of describing the performance of a line suite is to calculate four statistics from the set of $(\Oabu)_{\texttt{All\_NO\_S2}}$ (fiducial) versus $(\Oabu)_{\texttt{Suite\_Prior}}$ pairs generated by our Monte Carlo sampling over the observational sample whose distribution we show in the metallicity-metallicity plots. The first of these is the Pearson correlation coefficient, which measures on a scale from $-1$ to 1 how strongly the metallicity estimated with a particular line suite and weighting correlates with the fiducial metallicity. The second is the median orthogonal distance (MOD), which gives the median distance of points in the sample from the 1-to-1 line in the metallicity-metallicity plane (in units of dex), with positive values indicating that a given suite / weighting combination overestimates the metallicity relative to our fiducial case, and negative values indicating a tendency to underestimate the metallicity. Our third statistic is the root-mean-square orthogonal distance (RMSOD, again in units of dex), which quantifies the typical scatter in the metallicity derived using the proposed diagnostic relative to the fiducial one.\footnote{To see how the RMSOD differs from the Pearson correlation and MOD, and why it is a useful complement to them, consider a hypothetical metallicity diagnostic that produces a distribution of galaxies in the metallicity-metallicity plane that is very tight, and that matches the fiducial metallicity well at intermediate metallicities, but has a slope significantly shallower than unity and therefore underestimates high metallicities and overestimates low ones. For this case we would have a high Pearson correlation (since the distribution is tight) and possibly also a low MOD (since the over- and underestimates at the two ends balance to give a small median distance), but we would still have a large RMSOD.} The fourth and final statistic is the valid fraction, which is the fraction of output metallicity estimates using a particular line suite and weighting that lie within the bounds over which the metallicity estimators we use are valid, which runs from $\Oabu = 7.75$ to $9.252$. Inferences that land at these two boundaries are untrustworthy and must be discarded, so ideally the valid fraction should be close to 100\%.

\subsection{The Blue suite}
\label{sec:blue}

Our baseline model for the Blue suite, which we denote as \texttt{Blue}, uses as input to \nebbayes~all the lines for this suite listed in \autoref{sec:ins_z} with no line ratio prior setting, i.e., without applying any additional weights to the outcome based on particular line ratios. We also consider all possible priors that we could apply based on line ratios computable from the Blue suite, which we list in \autoref{tab:comb_names}. \texttt{Blue\_O2} uses the ratio [O~\textsc{ii}]$~\lambda 3729/$[O~\textsc{ii}]$~\lambda 3726$ as a prior to constrain the gas pressure \citep{kewley2019a};  \texttt{Blue\_O3\_O2}, \texttt{Blue\_O2\_O2}, and \texttt{Blue\_R23\_O2} use [O~\textsc{iii}]$/$H$\beta$, [O~\textsc{ii}]$/$H$\beta$, and $R_{23}$, respectively, to constrain the metallicity in addition to the pressure constraint from [O~\textsc{ii}]$~\lambda 3729/$[O~\textsc{ii}]$~\lambda 3726$ \citep{kewley2019b}.

A complication that arises with this line suite is that, due to the inclusion of both the [O~\textsc{ii}] doublet and the [O~\textsc{iii}], the posterior metallicity distributions of all combinations shows a behaviour similar to that of the traditional $R_{23}$ diagnostic, which has two branches on either side of a central metallicity $\Oabu = 8.4$ \citep{mcgaugh1991}. In our Bayesian framework, this manifests as the posterior probability distribution being bimodal, with two distinct peaks at low and high metallicity corresponding to the two branches. This creates two problems. First, standard summary statistics such as mode and mean are not useful descriptors for distributions of this shape. Second, this behaviour introduces a large discrepancy in the results between the Blue suite and the full optical line suite (which does not suffer from bimodal degeneracy), particularly for AGN hosts, that is somewhat artificial -- in practice the low-metallicity branch is implausible for an massive AGN in the local universe. To alleviate these problems, we do likewise by adopting as a prior that $\Oabu > 8.2$ and $>8.4$ for BPT-SF and AGN-host galaxies, respectively; this has the effect of suppressing the low-metallicity branch, and forcing our posteriors onto the upper branch. These values are the saddles of the overall posterior distributions for BPT-SF and AGN-host galaxies if we do not impose this prior. Moreover, the prior for BPT-SF targets is consistent with the point made by \citet{tremonti2004} that due to selection effects there are few SF galaxies with $\Oabu < 8.4$ in SDSS DR7 sample. Given that the metallicity estimated by \citeauthor{tremonti2004} is generally about 0.2 dex larger than the values returned by our Bayesian methods \citep{li2024}, 8.2 is thus a reasonable lower boundary for BPT-SF galaxies. Similarly, \cite{thomas2019} find that most SDSS AGN-host galaxies have $\Oabu > 8.4$, and thus imposing this as a prior to avoid the bimodal behaviour of the oxygen diagnostic lines is reasonable. However, this complication of selecting an appropriate branch remains an issue for purposes of using these diagnostics at higher redshift, where one may expect to see an increase in the population of low-metallicity galaxies based on previous work \citep[e.g.,][]{zahid2014}.\footnote{An alternative approach one might consider to break the metallicity degeneracy between the two branches is to use the ratio [O~\textsc{iii}]/[O~\textsc{ii}], which is monotonic in metallicity. Unfortunately this approach does not work because this ratio is significantly more sensitive to the ionization parameter and the hardness of the ionizing spectrum than it is to metallicity (e.g., see Figure 7 of \citealt{kewley2019b}). As a result, even small uncertainties in these parameters, which we cannot avoid given our limited line suites, preclude use of [O~\textsc{iii}]/[O~\textsc{ii}] as a means of breaking the $R_{23}$ branch degeneracy.}

We now compare the metallicity from the Blue suite with all its possible line ratio weightings (\autoref{tab:comb_names}) with the fiducial metallicites derived using the \texttt{All\_NO\_S2} suite. \autoref{tab:statistics_sf} and \autoref{tab:statistics_agn} report each of our four statistics separately for the BPT-SF and AGN-host galaxies, respectively, and we show the distribution of our sample in the metallicity-metallicity plane for all candidate weightings in \autoref{fig:sf_blue} and \autoref{fig:agn_blue} for BPT-SF and AGN-hosts galaxies, respectively. 

For BPT-SF galaxies, all the possible weightings using the Blue suite show similar features, with a median Pearson correlation coefficient of 0.879, median MOD of $0.073$, median RMSOD of 0.073, and median valid fraction of 98.2\%. The metallicity estimates from this line suite are in all cases strongly correlated with the fiducial metallicity, with a slight offset to higher metallicity ($\approx 0.06$ dex). The best combination, \texttt{Blue}, which we mark with an asterisk in \autoref{tab:statistics_sf}, shows the greatest Pearson correlation coefficient, smallest MOD and RMSOD, and greatest valid fraction with only $\sim$ 0.7\% of galaxies removed due to reaching the model boundaries. We show the metallicity-metallicity plot for this best-performing case in the upper left panel of \autoref{fig:performance}. As shown in this figure, at $\Oabu > 8.8$, the metallicity estimates from \texttt{Blue} generally follow the fiducial metallicity, but with an overall $\sim$ 0.06 dex offset to higher metallicity. At $\Oabu < 8.8$, the estimates start to deviate from the 1-to-1 relation, reaching a maximum offset $\sim$ 0.3 dex around $\Oabu = 8.4$. However, due to the small scatter, it is easy to fit two straight lines intersecting at $\Oabu = 8.8$ to remove this discrepancy.

Metallicity estimation for AGN hosts has larger uncertainties than for BPT-SF galaxies, as reflected by the generally smaller Pearson correlation coefficients in \autoref{tab:statistics_agn}. The different weightings we consider all provide similar performance, with median Pearson correlation coefficient of 0.525, median MOD of $-0.053$, median RMSOD of 0.125, and median valid fraction of 80.0\%. The combination \texttt{Blue} that works best for BPT-SF galaxies also shows the largest Pearson correlation coefficient and the smallest MOD and RMSOD, but the advantage is much smaller than for BPT-SF galaxies. We nonetheless consider it the best available combination due to this small advantage, and mark it with an asterisk in \autoref{tab:statistics_agn}. We show this distribution of sample galaxies in the metallicity-metallicity plane for this indicator in the lower left panel in \autoref{fig:performance}. We see that the majority of the sample roughly follow the 1-to-1 relation, but with a pronounced spread to low metallicity.

We can understand this spread as follows. As pointed out above, the $R_{23}$ metallicity diagnostic has two branches. \cite{kewley2019b} show that at low gas pressure $\log (P/k)$, the $R_{23}$ ratio at higher metallicity is smaller on the lower branch, while at high gas pressure, the ratio is similar on the two branches, which enhances the degeneracy of the diagnostics. Values of $\log (P/k)> 6.5$, the pressure at which this enhanced degeneracy appears, are uncommon in star-forming galaxies but relatively normal to find in NLRs \citep{kaddad2018}, driving an increase in the scatter of metallicities for AGN-host galaxies.

Another noteworthy feature of the lower left panel of \autoref{fig:performance} is the stripes at specific metallicities, indicating either an over-density (e.g. the stripes at $\Oabu < 8.8$) or an under-density (e.g. the stripes at $\Oabu > 8.8$). These stripes also appear in the work of \citet{tremonti2004}, and can similarly be seen in Figure 1 in \cite{lara2010}. We show in \aref{app:stripes} that these stripes are an artefact caused by the sparse sampling of the underling grid of photoionisation models, and could be improved in future work using a finer grid.

\subsection{The LRed suite}
\label{sec:lred}

As with the Blue suite, \autoref{tab:comb_names} shows all the emission-line ratio weightings available in the LRed suite. \texttt{LRed} does not weight by any line ratios, and is therefore analogous to \texttt{Blue}. \texttt{LRed\_S2} uses [S~\textsc{ii}]$~\lambda 6716/$[S~\textsc{ii}]$~\lambda 6731$ to constrain the gas pressure \citep{kewley2019a}. \texttt{LRed\_S2\_S2}, \texttt{LRed\_N2\_S2}, \texttt{LRed\_NS\_S2}, \texttt{LRed\_O3\_S2}, and \texttt{LRed\_O3N2\_S2} use [S~\textsc{ii}]$/$H$\alpha$, [N~\textsc{ii}]$/$H$\alpha$, [N~\textsc{ii}]$/$[S~\textsc{ii}], [O~\textsc{iii}]$/$H$\beta$, and ([O~\textsc{iii}]$/$H$\beta)/$([N~\textsc{ii}]$/$H$\alpha$)  to constrain the metallicity, respectively \citep{kewley2019b}, in addition to using [S~\textsc{ii}]$~\lambda 6716/$[S~\textsc{ii}]$~\lambda 6731$ to constrain the pressure.

For BPT-SF galaxies, all combinations show similar performance, with a tight 1-to-1 relation between the inferred and fiducial metallicity at $\Oabu > 8.8$ similar to produced by the Blue suite, but a large spread to low metallicity at $\Oabu < 8.8$ (as shown in \autoref{fig:sf_lred}). \autoref{tab:statistics_sf} shows that the median Pearson correlation coefficient of all combinations is 0.752, with a relatively small spread from one weighting to another; the exception is \texttt{LRed\_O3N2\_S2}, which shows a stronger correlation coefficient of 0.826. This is the best combination in this suite, and we show the metallicity-metallicity plot for it in the top-middle panel of \autoref{fig:performance}. It is obvious that this best Pearson coefficient is due to the smaller spread at $\Oabu < 8.8$ compared to the other available weighting schemes, most prominently \texttt{LRed\_N2\_S2} and \texttt{LRed\_NS\_S2}. Due to the spread to lower metallicity, the median MOD is negative, $-0.013$, with a median RMSOD of 0.118. The valid fractions are similar for all weightings, with a median of 99.3\%. Comparing with the Blue suite, the LRed suite has both a lower Pearson correlation coefficient and a larger RMSOD. The spread to low metallicity at $\Oabu < 8.8$ is primarily responsible for this worse performance. Due to this large scatter for metal-poor galaxies, and given that most BPT-SF galaxies at high redshift have $\Oabu < 8.8$ \citep{zahid2013}, we do not recommend using the LRed suite for BPT-SF galaxies if other suites are available.

In contrast, for AGN-host galaxies, the performance of LRed is competitive with the Blue suite. \autoref{fig:agn_lred} shows that different weightings all produce similar results, with the exception of \texttt{LRed\_O3N2\_S2}, for which most galaxies concentrate on several stripes, that, as explained in \aref{app:stripes}, are the result of limited sampling in our \mappings~grid. Inferences based on the LRed suite generally yield metallicities smaller than the fiducial estimate. The tightest relation between the metallicities estimated from the reduced line suite and the fiducial values is for \texttt{LRed\_NS\_S2}, which is the best combination for AGN-host galaxies and is plotted in the bottom-middle panel of \autoref{fig:performance}; however, it performs only marginally better than several other weighting options. As quantified in \autoref{tab:statistics_agn}, the median Pearson correlation coefficient is 0.504, and \texttt{LRed\_NS\_S2}, the best combination, shows the highest Pearson correlation of 0.556, which is better than \texttt{Blue} in the Blue suite. The median MOD is negative, $-0.04$, while the median RMSOD is 0.101. This is consistent with our impression from \autoref{fig:agn_lred}. The median valid fraction is 87.9\%. The larger Pearson correlation coefficient, smaller RMSOD, and larger valid fraction of LRed compared to Blue make LRed the best emission-line suite for AGN-host galaxies.

\subsection{The SRed suite}
\label{sec:sred}

SRed is a subset of LRed, and thus offers a subset of the weighting choices available for SRed; we list these in \autoref{tab:comb_names} as well. In analogy with the other suites, \texttt{SRed} does not use any line ratio weightings, \texttt{SRed\_S2} uses [S~\textsc{ii}]$~\lambda 6716/$[S~\textsc{ii}]$~\lambda 6731$ to constrain the gas pressure \citep{kewley2019a}, and \texttt{SRed\_S2\_S2}, \texttt{SRed\_N2\_S2}, and \texttt{SRed\_NS\_S2} add [S~\textsc{ii}]$/$H$\alpha$, [N~\textsc{ii}]$/$H$\alpha$, and [N~\textsc{ii}]$/$[S~\textsc{ii}], respectively, as prior constraints on the metallicity in addition to the pressure constraint \citep{kewley2019b}. As with other line suites, we report the performance results for these options for BPT-SF galaxies and AGN-host in \autoref{tab:statistics_sf} in \autoref{tab:statistics_agn}, and show the metallicity-metallicity distributions in \autoref{fig:sf_sred} and \autoref{fig:agn_sred}, respectively. 

For BPT-SF galaxies, different weighting options in SRed show divergent performance. \texttt{SRed\_S2\_S2} has little capability to constrain metallicity as the relation is basically flat and the Pearson correlation coefficient is 0.315. \texttt{SRed} and \texttt{SRed\_S2} are similar, with a bias to systematically lower metallicity than the fiducial case (MOD $-0.134$ for \texttt{SRed} and $-0.140$ for \texttt{SRed\_S2}), while \texttt{SRed\_N2\_S2} has slightly better performance with a positive MOD of 0.06. The highest Pearson correlation is for \texttt{SRed\_NS\_S2}, which we show in the top-right panel of \autoref{fig:performance}. At $\Oabu > 8.8$ where both Blue and LRed perform well, \texttt{SRed\_NS\_S2} generally follows the 1-to-1 relation with a slightly larger scatter than the other two suites, consistent with this case having a slightly larger RMSOD and smaller Pearson correlation than \texttt{Blue}. In addition, there is a clustering at $\Oabu = 9.096$ as indicated by the stripe in the top-right panel in \autoref{fig:performance} and by the peak of the distribution in the top-right panel in \autoref{fig:stripes}, giving rise to a large uncertainty for the metallicity estimate around that value. However, the performance of this suite at $\Oabu < 8.8$ follows the 1-to-1 relation better than the LRed suite, and avoids the large spread to lower metallicity seen in LRed (see the top-middle and top-right panel in \autoref{fig:performance} for comparison). This is why both the Pearson correlation coefficient and the RMSOD of \texttt{SRed\_NS\_S2} are better than for \texttt{LRed\_O3N2\_S2}, and this makes the SRed a better suite for the metallicity estimation at this range. This is a somewhat surprising result, since SRed is a subset of LRed, and thus we might expect LRed to perform better in all cases; it is clear that this expectation is incorrect, and that the addition of H$\beta$ and [O~\textsc{iii}] (the lines present in LRed and absent in SRed) do not by themselves improve performance. This may be related to the bimodality of the [O~\textsc{iii}] line intensity discussed in \autoref{sec:blue}.

For AGN-host galaxies we also find that different weighting combinations yield very different performance as illustrated in \autoref{fig:agn_sred} and quantified in \autoref{tab:statistics_agn}. As with BPT-SF galaxies, the \texttt{SRed\_S2\_S2} combination offers very little constraint on the metallicity, with a Pearson correlation coefficient of only 0.375. \texttt{SRed\_N2\_S2} and \texttt{SRed\_NS\_S2} also perform poorly, with many metallicities reaching the upper boundary of our model grids, yielding valid fractions of only 35.9\% and 20.4\%, respectively; this failure is visible in the bottom panels of \autoref{fig:agn_sred}. Finally, \texttt{SRed} and \texttt{SRed\_S2} have similar performances; we choose \texttt{SRed} as the best combination in this suite due to its greater Pearson correlation coefficient and substantially smaller MOD, and plot it in the bottom-right panel in \autoref{fig:performance}. However, this diagnostic shows pronounced clustering of galaxies at $\Oabu = 8.997$ and $\Oabu = 9.180$, reflected by the two stripes in the bottom-right panel in \autoref{fig:performance} and two peaks in the bottom-right panel in \autoref{fig:stripes}. This clustering gives rise to large uncertainties for the metallicity estimates and makes \texttt{SRed} substantially worse than the best-performing weightings for the Blue and LRed suites. We therefore do not recommend use of the SRed suite for AGN host galaxies if other suites are available.

\subsection{Summary of test results and recommendations for future surveys}
\label{sec:sum_performance}

We are now in a position to make some general recommendations about strategies for future surveys, and in particular about which line suites to use given the galaxy population one wishes to target. In making these recommendations, it is important to keep in mind that no matter what line suite one chooses performance will be worse for AGN-host galaxies, which is understandable due to the diversity of AGN ionizing spectra and larger number of parameters required to describe the NLRs. With this caveat in mind, we can recommend the following:

\begin{itemize}
    \item The Blue suite (using the \texttt{Blue} weighting, i.e., unweighted) performs quite well overall, particularly for BPT-SF galaxies, and it is available to the highest redshifts (\autoref{fig:suites_redshift}). We therefore recommend it with one significant caution, which is that the posterior distributions it produces are bimodal, and that this bimodality is particularly severe for AGN hosts. It should therefore only be used on galaxy populations where one is comfortable imposing a prior to force the posterior onto one of the two branches. However, as there are more faint and metal poor galaxies at early universe, this prior excludes these galaxies and reduces the sample completeness at higher redshift. If such completeness is required, then the other two suites or the full-line suite should be considered.
    \item The LRed suite shows the best performance for AGN hosts, while SRed performs very poorly for them and is not recommended. For BPT-SF galaxies, at $\Oabu > 8.8$ LRed shows a strong and tight correlation with the fiducial estimator, while the same is true of SRed at $\Oabu < 8.8$. A mixed usage of LRed and SRed is recommended, depending on the expected metallicity range of the target population.
    \item If the red suites are used (SRed or LRed), based on the point above, a useful strategy would be to observe targets in the windows allowed by LRed, so that it is possible to use both LRed, which is the best suite for AGN hosts, and a mixture of LRed and SRed for BPT-SF galaxies. With this strategy, AGN hosts and non-AGN galaxies can be directly compared.  
\end{itemize}

\section{Signal-to-noise ratio effects}
\label{sec:snr}

\begin{table*} 
 \centering
 \renewcommand{\arraystretch}{1.3}
 \begin{threeparttable}
  \caption{Same as \autoref{tab:statistics_sf}, but now showing performance statistics comparing metallicity estimates with the best-performing weighting combination for each line suite (marked with asterisks in \autoref{tab:statistics_sf}) but reduced SNRs to estimates using data with the native SDSS SNR. The Blue and LRed suites use SNR$_{\mathrm{H}\beta}=25$ as the reference sample, and compare the performances with samples where this is reduced to 15, 10, 5, as indicated by the notations \texttt{H$\beta$\_15}, \texttt{H$\beta$\_10}, and \texttt{H$\beta$\_5}. The SRed suite uses SNR$_{\mathrm{H}\alpha}=40$ as the reference sample, which is compared to cases where the SNR is reduces to 30, 20, and 10.}
  \vspace{1mm}
  \label{tab:snr_sf}

  \begin{tabular}{lcccc}
   \hline
   Name & Pearson coefficient & Median orthogonal distance (dex) & RMS orthogonal distance (dex) & Valid fraction \\
   \hline
   \texttt{Blue\_H$\beta$\_15} & 0.924 & 0.000 & 0.034 & 99.9\% \\
   \texttt{Blue\_H$\beta$\_10} & 0.877 & 0.000 & 0.048 & 99.6\% \\
   \texttt{Blue\_H$\beta$\_5} & 0.708 & 0.007 & 0.091 & 98.8\% \\
   \hline
   \texttt{LRed\_O3N2\_S2\_H$\beta$\_15} & 0.896 & 0.007 & 0.078 & 100\% \\
   \texttt{LRed\_O3N2\_S2\_H$\beta$\_10} & 0.849 & 0.007 & 0.099 & 99.8\% \\
   \texttt{LRed\_O3N2\_S2\_H$\beta$\_5} & 0.713 & 0.020 & 0.149 & 98.3\% \\
   \hline
   \texttt{SRed\_NS\_S2\_H$\alpha$\_30} & 0.933 & 0.007 & 0.055 & 99.8\% \\
   \texttt{SRed\_NS\_S2\_H$\alpha$\_20} & 0.903 & 0.013 & 0.073 & 99.6\% \\
   \texttt{SRed\_NS\_S2\_H$\alpha$\_10} & 0.823 & 0.000 & 0.112 & 99.6\% \\
   \hline   
  \end{tabular} 
 \end{threeparttable}
\end{table*}

\begin{table*} 
 \centering
 \renewcommand{\arraystretch}{1.3}
 \begin{threeparttable}
  \caption{The same as \autoref{tab:snr_sf}, but for AGN-host galaxies.}
  \vspace{1mm}
  \label{tab:snr_agn}

  \begin{tabular}{lcccc}
   \hline
   Name & Pearson coefficient & Median orthogonal distance (dex) & RMS orthogonal distance (dex) & Valid fraction \\
   \hline
   \texttt{Blue\_H$\beta$\_15} & 0.823 & 0.000 & 0.072 & 87.7\% \\
   \texttt{Blue\_H$\beta$\_10} & 0.717 & 0.007 & 0.095 & 80.4\% \\
   \texttt{Blue\_H$\beta$\_5} & 0.416 & 0.000 & 0.156 & 64.0\% \\
   \hline
   \texttt{LRed\_NS\_S2\_H$\beta$\_15} & 0.923 & 0.000 & 0.056 & 97.6\% \\
   \texttt{LRed\_NS\_S2\_H$\beta$\_10} & 0.884 & 0.000 & 0.068 & 96.2\% \\
   \texttt{LRed\_NS\_S2\_H$\beta$\_5} & 0.783 & 0.000 & 0.094 & 92.7\% \\
   \hline
   \texttt{SRed\_H$\alpha$\_30} & 0.925 & 0.000 & 0.052 & 98.3\% \\
   \texttt{SRed\_H$\alpha$\_20} & 0.895 & 0.000 & 0.062 & 96.5\% \\
   \texttt{SRed\_H$\alpha$\_10} & 0.796 & 0.000 & 0.088 & 91.6\% \\
   \hline   
  \end{tabular} 
 \end{threeparttable}
\end{table*}

We next investigate how the performance of these different line suites changes as a function of signal-to-noise ratio (SNR). We focus on the best combinations in each emission-line suite for both BPT-SF and AGN-host galaxies, as plotted in \autoref{fig:performance} and marked with asterisks in \autoref{tab:statistics_sf} and \autoref{tab:statistics_agn}. Our goal is to provide guidance that can be used to design future surveys aimed at deriving metallicity using the instruments considered here (or indeed any with similar wavelength divisions).

To accomplish this, our strategy is to select galaxies from our parent sample with high SNR, then degrade the SNR, and and examine how the derived metallicities change with respect to the original, high SNR data. As mentioned in \autoref{sec:sample}, to avoid introducing a sampling bias, we compute SNRs only for the Balmer lines, using H$\beta$ when it is available (as it is in the Blue and LRed suites) and H$\alpha$ otherwise (for the SRed suite). We select galaxies with SNR$_{\mathrm{H}\beta} = 25$, with this value chosen to balance sample size with having a starting SNR high enough to allow us to explore a wide range of signal degradation. For the SRed suite, we set corresponding SNR$_{\mathrm{H}\alpha} = 40$ based on the intrinsic Balmer decrement. The resulting sample sizes for BPT-SF galaxies are 5173 based on H$\beta$, and 6386 based on H$\alpha$, while for AGN-host galaxies our SNR cut leaves 336 based on H$\beta$, and 436 based on H$\alpha$.

To test how performance depends on SNR, for each galaxy in our sample we scale the flux errors for each line by a common factor to achieve target SNRs of SNR$_{\mathrm{H}\beta}=15, 10, 5$ for the Blue and LRed suites, and SNR$_{\mathrm{H}\alpha}=30, 20, 10$ for the SRed suite, giving us four versions of the line fluxes for each galaxy -- one at the original SDSS SNR, and one at each at the lower SNR levels. As in \autoref{sec:nb}, we randomly generate twenty realizations for each galaxy at each SNR level by drawing a flux for each line from its uncertainty distribution.\footnote{Note that as we elaborate in \autoref{sec:nb}, we discard and redraw realizations where any of the strong-line fluxes are negative. Such realizations rise to a maximum of $\sim 8\%$ of total cases for the lowest SNR experiments (SNR$_{\mathrm{H}\alpha}=10$ and SNR$_{\mathrm{H}\beta}=5$). We test whether discarding these cases introduces any significant bias by trying the alternative procedure of keeping those discarded realizations but assigning a positive value equal to 5 per cent of the corresponding flux
error for any line with negative flux. We find that this leads to only very small changes from the outcome of our fiducial procedure, and thus conclude that the bias is not large.} We then estimate the metallicities for each realization of each galaxy at each SNR level with the best weighting combinations for each line suite as determined in \autoref{sec:analysis}: \texttt{Blue} for both the Blue suite for both BPT-SF and AGN-host galaxies, \texttt{LRed\_O3N2\_S2} for BPT-SF and \texttt{LRed\_NS\_S2} for AGN-host galaxies in the LRed suite, and \texttt{SRed\_NS\_S2} for BPT-SF and \texttt{LRed} for AGN-host galaxies in the SRed suite. We evaluate the effects of SNR in these trials by comparing the metallicities derived at the full, native SNR with those derived using degraded SNRs using metallicity-metallicity plots together with the same statistics as used in \autoref{sec:analysis}, in this case taking the full-SNR result as our fiducial estimate to place on the horizontal axis.

We report the quantitative statistical results for each line suite and SNR ratio in \autoref{tab:snr_sf} for BPT-SF galaxies and in \autoref{tab:snr_agn} for AGN hosts, respectively. We also provide metallicity-metallicity plots for all these cases in \aref{app:SNR}. It is clear  and unsurprising that performance degrades as we lower the SNR, as reflected by the Pearson correlation coefficients and the RMSOD. We focus on the latter as the key parameter for this test, since can be roughly treated as the level of random uncertainty we can expect from observations at a given SNR. 

In general the scatters are $\lesssim 0.1$ dex except for the worst SNRs we consider (SNR$_{\mathrm{H}\beta} = 5$ or SNR$_{\mathrm{H}\alpha} = 10$), and we note that the Blue suite for AGN hosts and the LRed suite for BPT-SF galaxies appear to be the most sensitive to noise, while Blue for BPT-SF and SRed for AGN hosts are the least sensitive. A good rule of thumb based on our results is that surveys that seek to have random uncertainties in metallicity $\lesssim 0.1$ dex should target SNR$_{\mathrm{H}\beta} \geq 10$, but the exact level of accuracy needed must of course be designed with the science goals of the survey in mind. \autoref{tab:snr_sf} and \autoref{tab:snr_agn} can then be used to read off the SNR required for those accuracy goals to be achieved.

\section{Summary}
\label{sec:summary}

Measuring and comparing the metallicities, and the internal metallicity distributions, for non-AGN and AGN-host galaxies in the early universe is a promising avenue for understanding the role of AGN in regulating galaxy formation. Bayesian methods make such comparisons possible using self-consistent techniques in the Local Universe, but these require access to a large number of emission lines spread across the rest optical band, from the [O~\textsc{ii}]$~\lambda\lambda3727,29$ doublet on the blue end to the [S~\textsc{ii}]$~\lambda 6716/$[S~\textsc{ii}]$~\lambda 6731$ doublet on the red end. Beyond the local Universe, simultaneous measurements of all these lines will not be possible for the next generation of AO-assisted IFS instruments such as MAVIS, GMTIFS, and HARMONI. If we require all the optical lines to obtain reliable metallicity measurements that allow meaningful comparison of AGN-host and non-host galaxies, this means we will be forced into expensive repeated observations. This situation motivates us to explore how well we can measure metallicities with Bayesian techniques using only a subset of emission lines that can be observed simultaneously in a single-band measurement.

Based on the clustering of emission lines in the optical band, we identify three candidate emission line suites for such single-epoch observations, which we denote Blue, LRed, and SRed (\autoref{sec:z_range}). Within these suites there is at least one line ratio that is known to be sensitive to metallicity. We identify the redshift ranges over which each suite is accessible in a single-epoch observation with each observing band of next-generation instruments (\autoref{tab:redshift}, \autoref{fig:suites_redshift}), and then consider a range of weighting strategies in the Bayesian metallicity estimation code \textsc{NebulaBayes} to measure metallicities using only the lines accessible within each suite. We characterise the performance of these strategies by comparing them to fiducial metallicity estimates derive from all optical lines for a sample of nearby AGN-host and non-AGN galaxies drawn from the SDSS. 

We find that metallicity estimators for AGN hosts have larger uncertainties than for non-host galaxies across all suites. For the Blue suite, the best-performing is the \texttt{Blue} combination using no additional weighting for both AGN hosts and non-AGN galaxies. This suite overall performs well and allows access to the highest redshifts, with the caveat that this good performance is based on being able to adopt a prior to break the degeneracy between high- and low-metallicity branches of the traditional $R_{23}$ diagnostic. This prior reduces the sample completeness at higher redshift. If such completeness is required, then the other two suites or the full-line suite should be considered. For Red suites (LRed and SRed), we find the LRed is the best suite for AGN hosts, and a mixed usage of LRed and SRed is recommended for non-AGN galaxies, depending on the expected metallicity range of the target population (see \autoref{sec:sum_performance} for details). We therefore recommend observing targets in the windows allowed by LRed so that direct comparison between AGN hosts and non-AGN galaxies is possible. For the convenience, we provide full quantification of the performance of each possible strategy in \autoref{tab:statistics_sf} and \autoref{tab:statistics_agn}.

Finally, we examine the performance of each metallicity diagnostic as we degrade the signal-to-noise ratio (SNR) of the data. We find that for most line suites a SNR of 10 in the H$\beta$ line, or 20 in the H$\alpha$ line, is sufficient to reduce RMS metallicity errors induced by noise to $\lesssim 0.1$ dex, and we also identify particular line suites that are more and less sensitive to noise (\autoref{tab:snr_sf} and \autoref{tab:snr_agn}). These tabulations enable readers to identify a line suite and SNR target suited to their particular scientific goals, and thereby to calculate the exposure times required to achieve those goals with future instruments.

\textcolor{red}{}

This study can be extended by using the best-performing strategies for each line suite to investigate how loss of some of the optical lines available locally and finite SNR influence measurements of the metal distribution fields within galaxies, rather than simply the mean metallicity on which we have focused here. In particular, it is important to explore the ability of each line suite to recover the correlation lengths of galaxy metallicity fields, which \citet{li2025} demonstrate are particularly powerful tools for understanding the role of AGN in galactic ecology. This question can be investigated using simulated galaxies by extending the methods outlined by \citet{liz2024}, who focus on star-forming galaxies, to simulations including AGN.

\section*{Acknowledgements}

This study used computational resources provided by
the National Computational Infrastructure, which is supported by the Australian Government, through award jh2. MRK acknowledges support from the Australian Research Council through Laureate Fellowship FL220100020.

\section*{Data Availability}

The data underlying this article will be shared on reasonable request with the corresponding author.



\bibliographystyle{mnras}
\bibliography{Li24c} 



\appendix

\section{Metallicity-metallicity plots for all line weighting combinations}
\label{app:performance_all}

\begin{figure*}
    \resizebox{16cm}{!}{\includegraphics{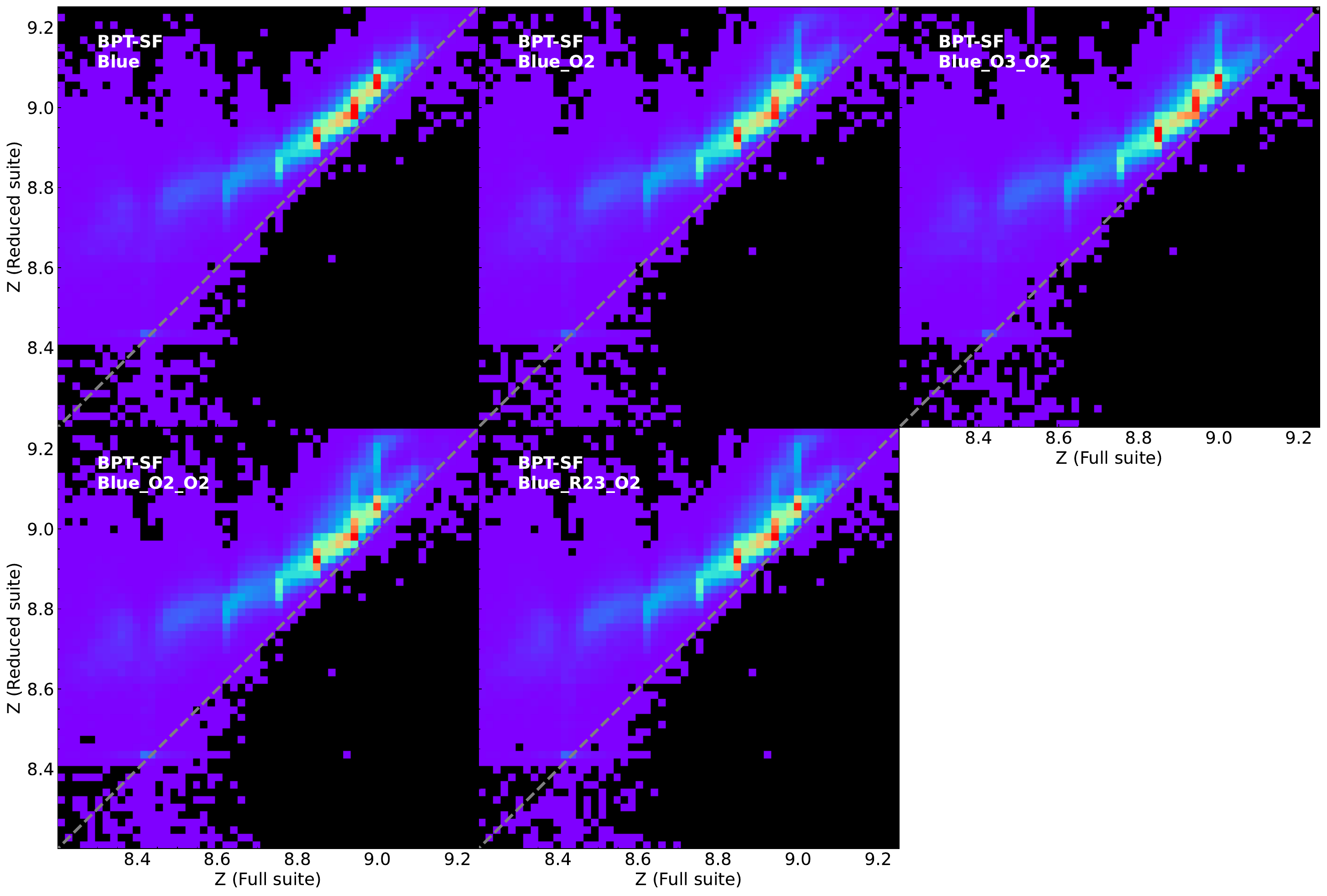}}
    \caption{The same as \autoref{fig:performance}, but shows the performances of all combinations of Blue suite for BPT-SF galaxies.}
    \label{fig:sf_blue}
\end{figure*}

\begin{figure*}
    \resizebox{16cm}{!}{\includegraphics{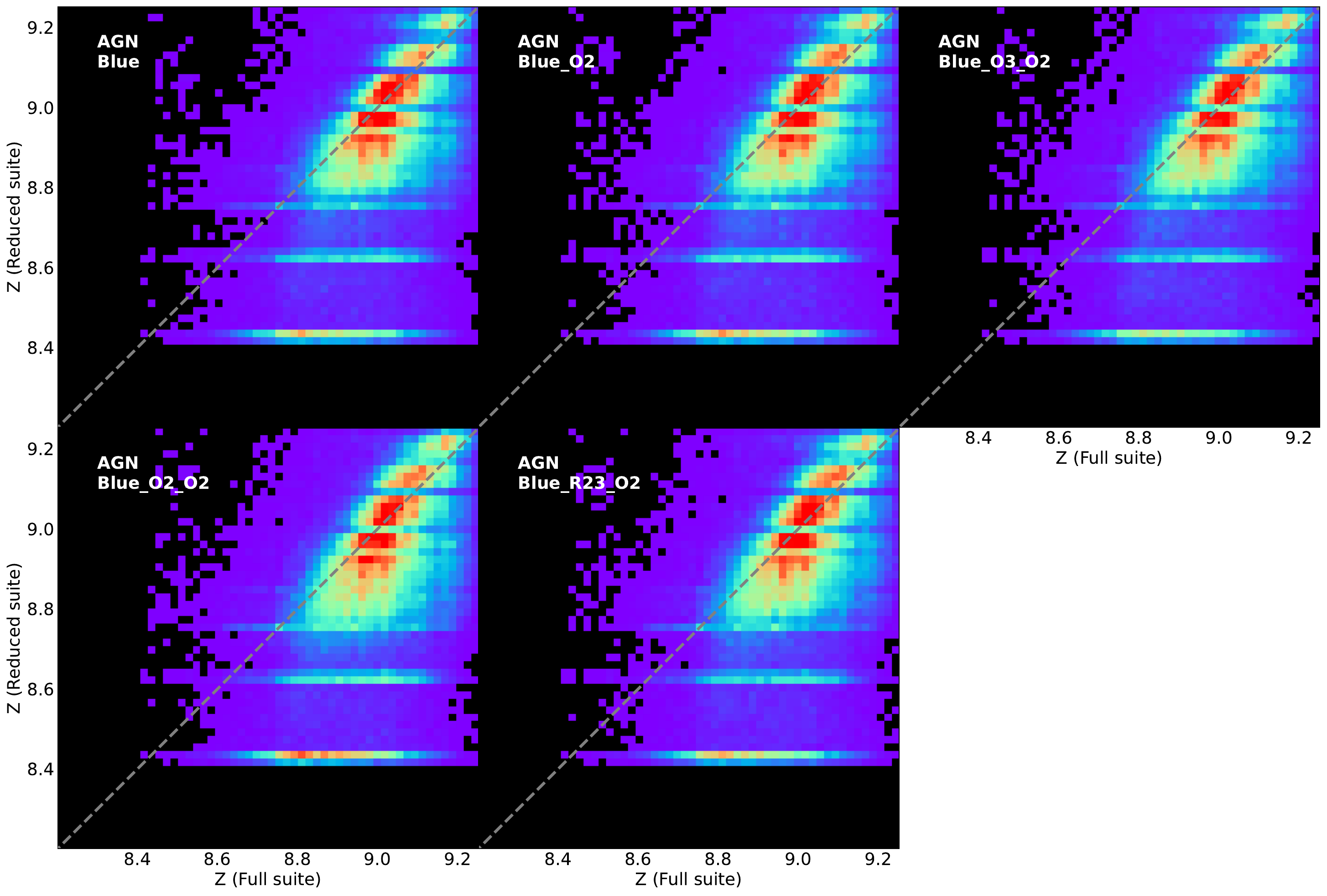}}
    \caption{The same as \autoref{fig:performance}, but shows the performances of all combinations of Blue suite for AGN-host galaxies.}
    \label{fig:agn_blue}
\end{figure*}

\begin{figure*}
    \resizebox{16cm}{!}{\includegraphics{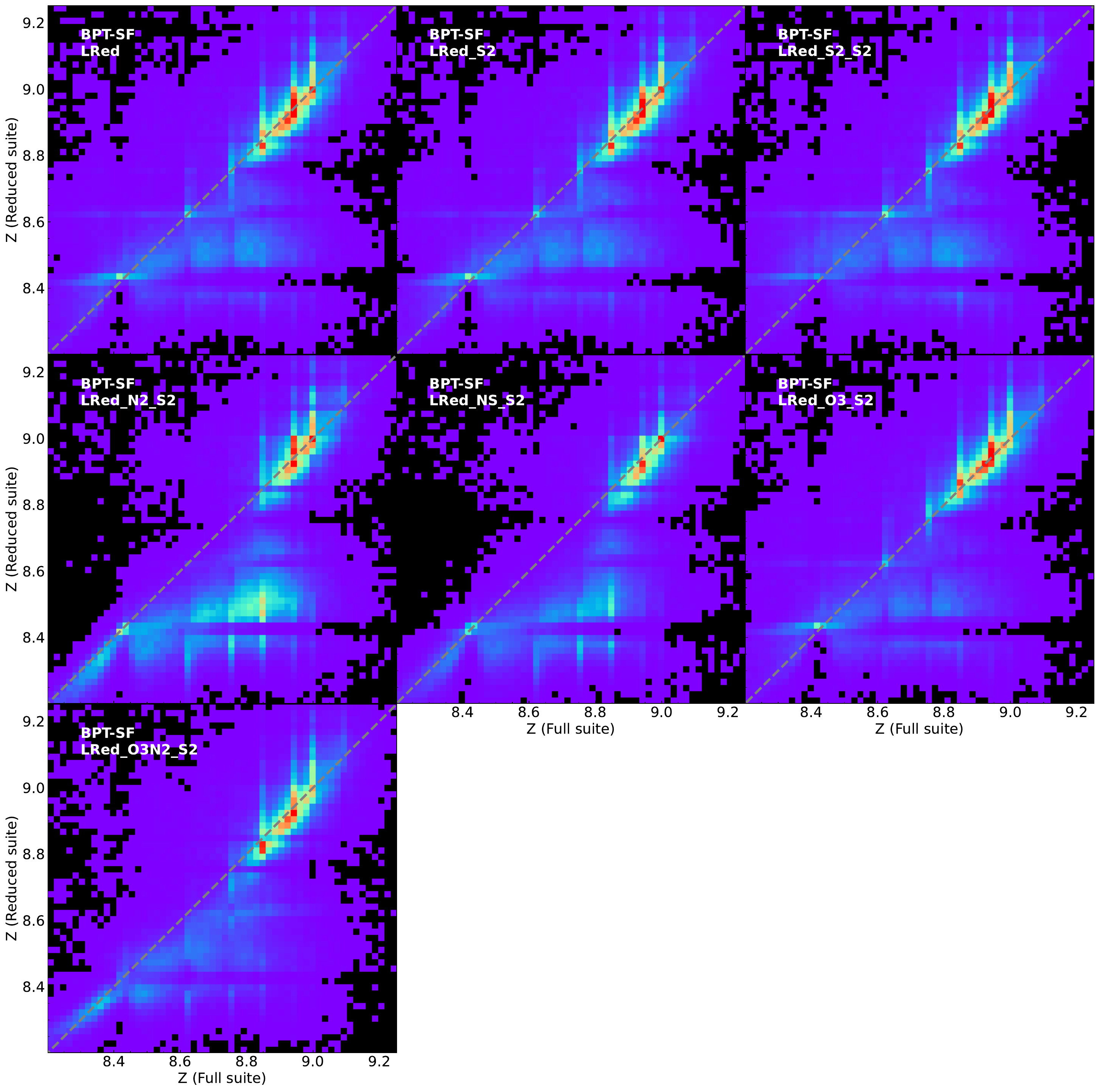}}
    \caption{The same as \autoref{fig:performance}, but shows the performances of all combinations of LRed suite for BPT-SF galaxies.}
    \label{fig:sf_lred}
\end{figure*}

\begin{figure*}
    \resizebox{16cm}{!}{\includegraphics{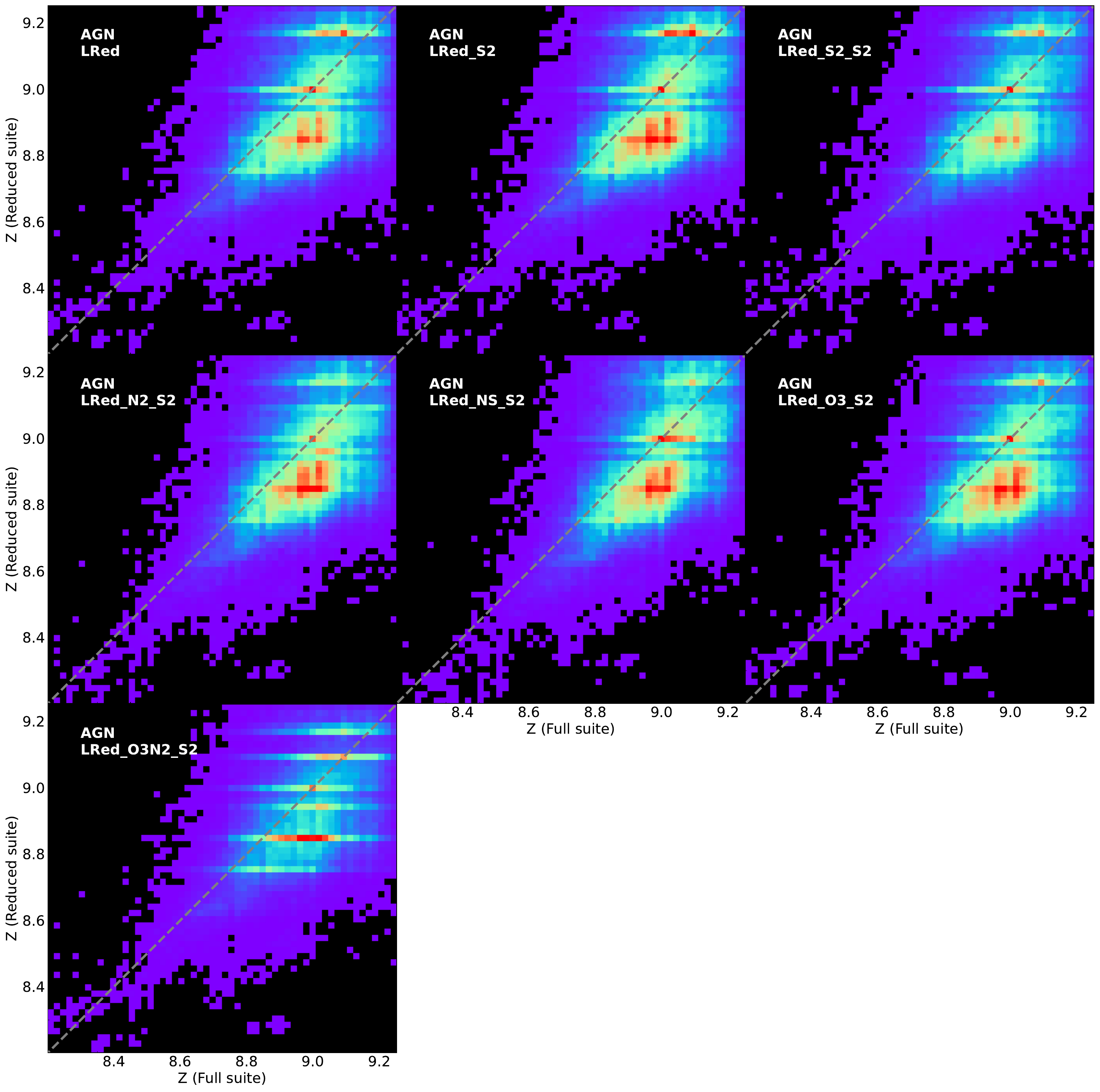}}
    \caption{The same as \autoref{fig:performance}, but shows the performances of all combinations of LRed suite for AGN-host galaxies.}
    \label{fig:agn_lred}
\end{figure*}

\begin{figure*}
    \resizebox{16cm}{!}{\includegraphics{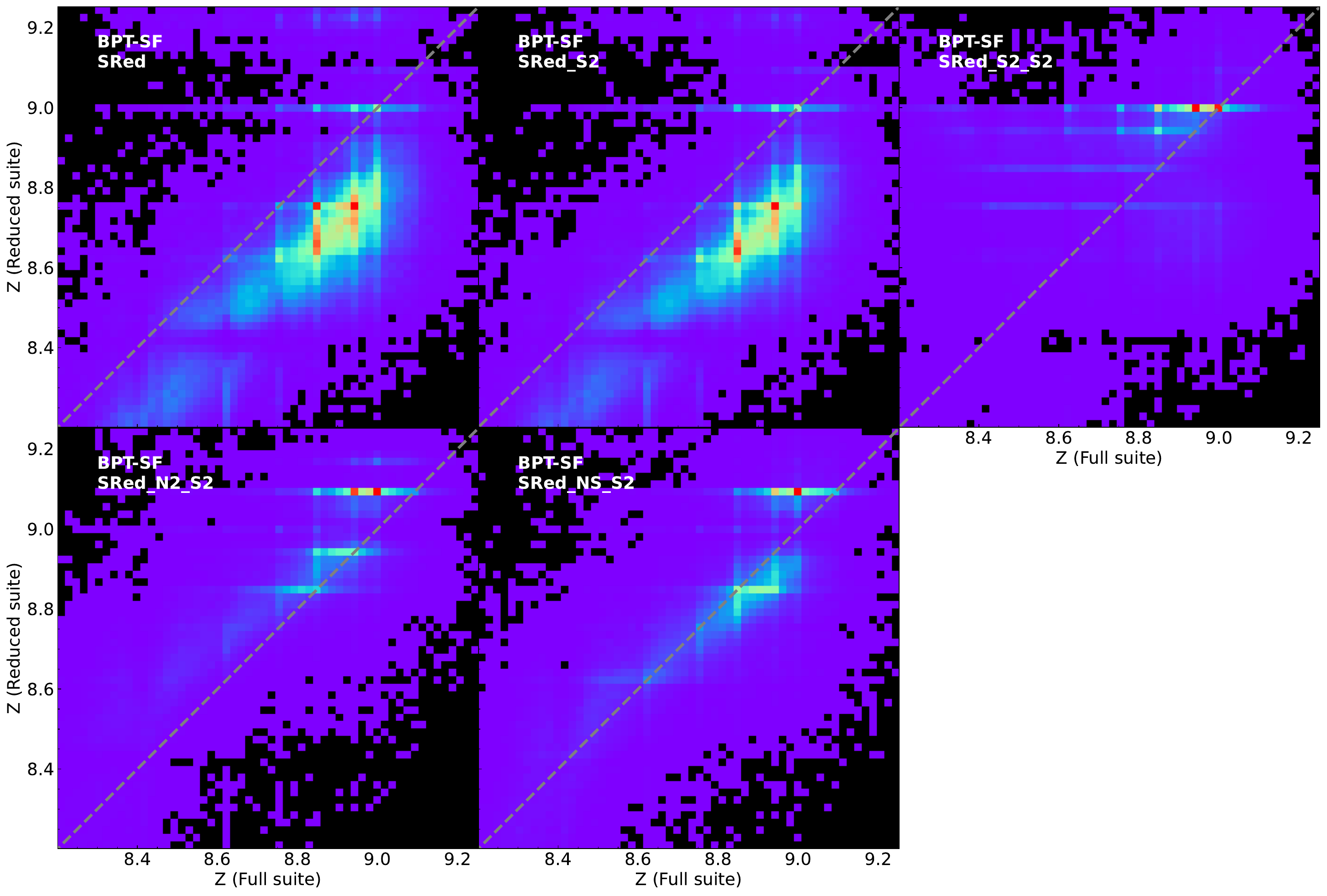}}
    \caption{The same as \autoref{fig:performance}, but shows the performances of all combinations of SRed suite for BPT-SF galaxies.}
    \label{fig:sf_sred}
\end{figure*}

\begin{figure*}
    \resizebox{16cm}{!}{\includegraphics{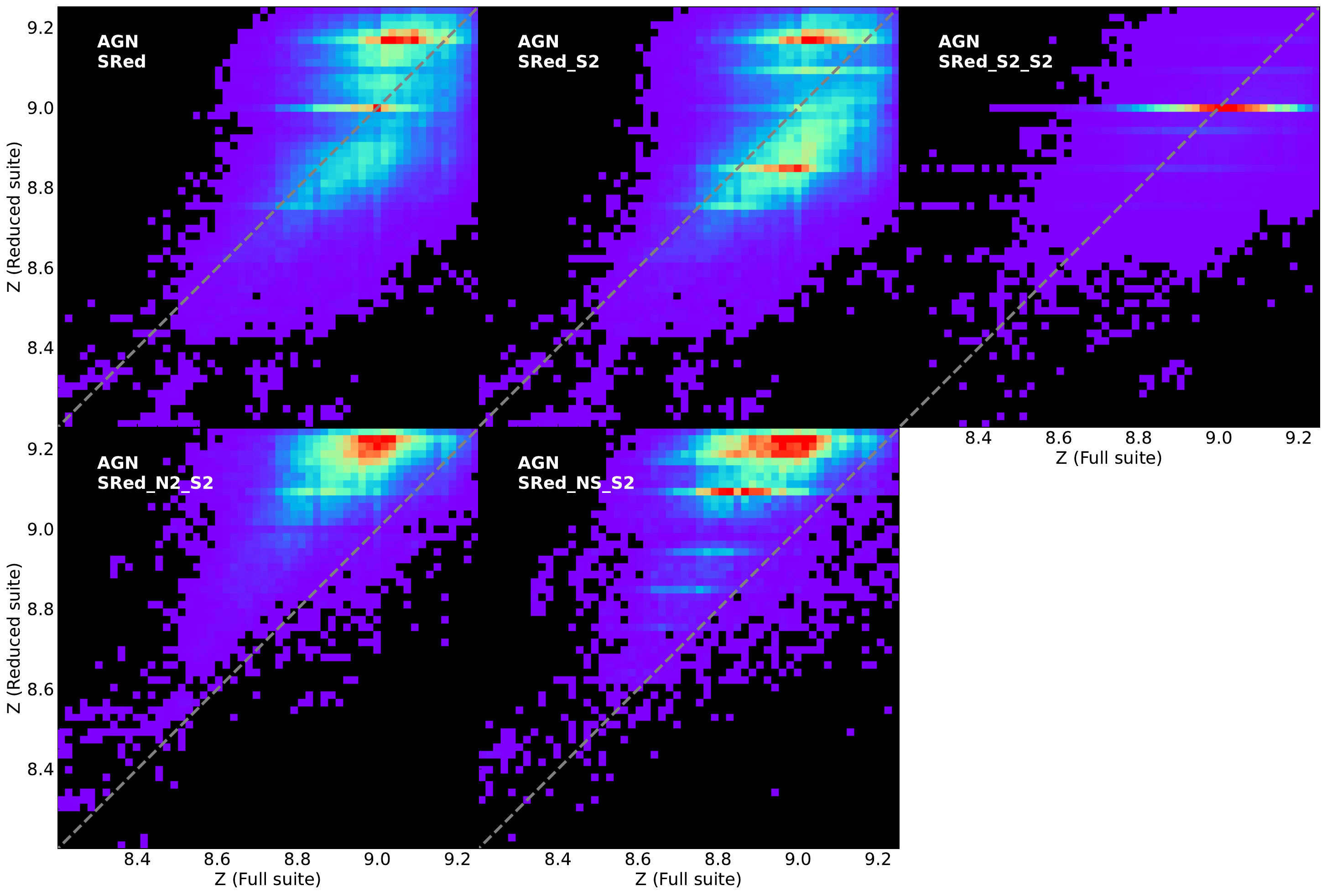}}
    \caption{The same as \autoref{fig:performance}, but shows the performances of all combinations of SRed suite for AGN-host galaxies.}
    \label{fig:agn_sred}
\end{figure*}

In \autoref{sec:analysis} we only show metallicity-metallicity plots for the best line weighting choice for each suite for BPT-SF and AGN hosts galaxies (\autoref{fig:performance}). Here for completeness we present plots for all the weighting choices we test: \autoref{fig:sf_blue} and \autoref{fig:agn_blue} show different line weighting options for the Blue suite for BPT-SF and AGN-host galaxies, respectively, while \autoref{fig:sf_lred}, \autoref{fig:agn_lred}, \autoref{fig:sf_sred}, and \autoref{fig:agn_sred} show the analogous plots for the LRed and SRed suites; panels are ordered in the same way as the statistics shown in \autoref{tab:statistics_sf} and \autoref{tab:statistics_agn}.

\section{The origin of striping in metallicity-metallicity plots}
\label{app:stripes}

\begin{figure*}
    \resizebox{16cm}{!}{\includegraphics{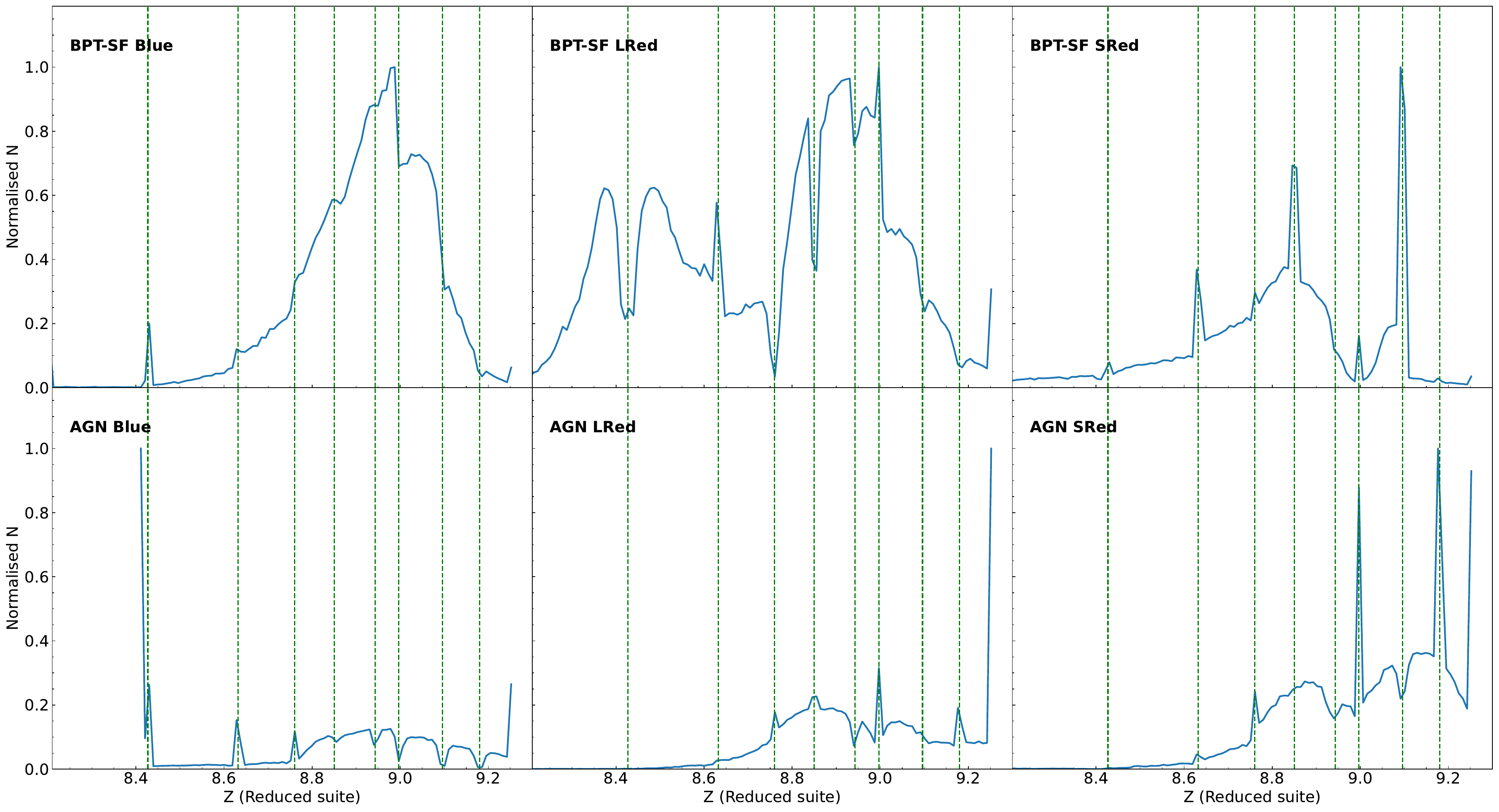}}
    \caption{The metallicity distributions of BPT-SF galaxies (top row) and AGN hosts (bottom row) estimated by best-performing weighting  of each emission-line suite (models marked with asterisks in \autoref{tab:statistics_sf} and \autoref{tab:statistics_agn}). Panels follow the same ordering as in \autoref{fig:performance}. The dashed lines indicate the values of metallicity on which the \mappings~grid used as input to \nebbayes~is computed. It is clear that the extrema in the metallicity distribution occur at these grid points. This shows that the stripes seen in \autoref{fig:performance} are due to the sparse sampling of the model grids.}
    \label{fig:stripes}
\end{figure*}

\autoref{fig:performance} shows the performances of the best weightings of each emission-line suite. However, most of these diagnostics show some degree of striping over at least part of the true metallicity range, which take the form of either an over-density or an under-density of galaxies. To understand what causes these stripes, in \autoref{fig:stripes} we show the distribution of metallicities for all BPT-SF galaxies (top row) and AGN hosts (bottom row) estimated using the best-performing weighting of each emission-line suite, i.e., using the same diagnostics as shown in \autoref{fig:performance}. The clear peaks and ditches are observed in all six panels corresponding to the clustering and deficiencies in \autoref{fig:performance}. In this figure, however, we also indicated by the dashed vertical lines the metallicities corresponding to our grid of \mappings~models (see the Appendix B3 of \citet{li2024} for details). We see that all the peaks and ditches occur at values corresponding to the grid points.

As explained in \citet{li2024}, we determine line values at metallicities intermediate between the grid point values by linear interpolation, but the clustering of the data indicates that our grid is likely not fine enough to capture the behaviour around extrema in the values of line ratios. These extrema correspond to local best- or worst-fits, and our procedure of assigning the mode of the posterior PDF, i.e., the single best-fitting model, as the ``true metallicity'' then responds to these extrema by either clustering points on them or avoiding them. This analysis suggests that a finer grid may find higher or lower values to match the data and reduce the striping, but at the price of more computing power.

As noted in the main text, similar stripes can also be observed in Figure 1 in \citet{lara2010}, which shows the metallicity estimated by \citet{tremonti2004} who also adopt Bayesian methods. This suggests that striping is a systematic limitation of at least some Bayesian methods when used on with a moderate-resolution grid.

\section{Metallicity-metallicity plots for the signal-to-noise ratio tests}
\label{app:SNR}

\begin{figure*}
    \resizebox{16cm}{!}{\includegraphics{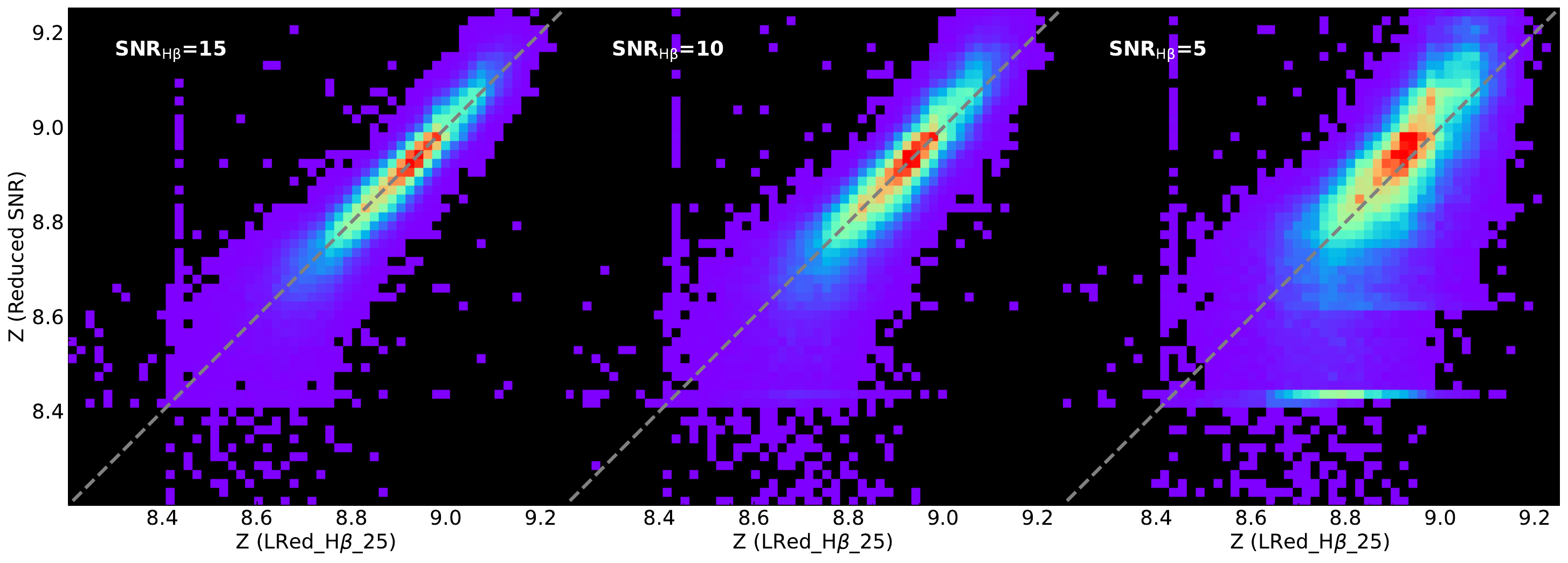}}
    \caption{The SNR tests for \texttt{Blue} combination in Blue suite for BPT-SF galaxies. The reference SNR$_{\mathrm{H}\beta}=25$, and we test 15, 10, and 5 in three panels. The counts of each bin are normalised to the maximum of each panel. From purple to red, the histograms show the range from 0 to 0.9.}
    \label{fig:snr_sf_blue}
\end{figure*}

\begin{figure*}
    \resizebox{16cm}{!}{\includegraphics{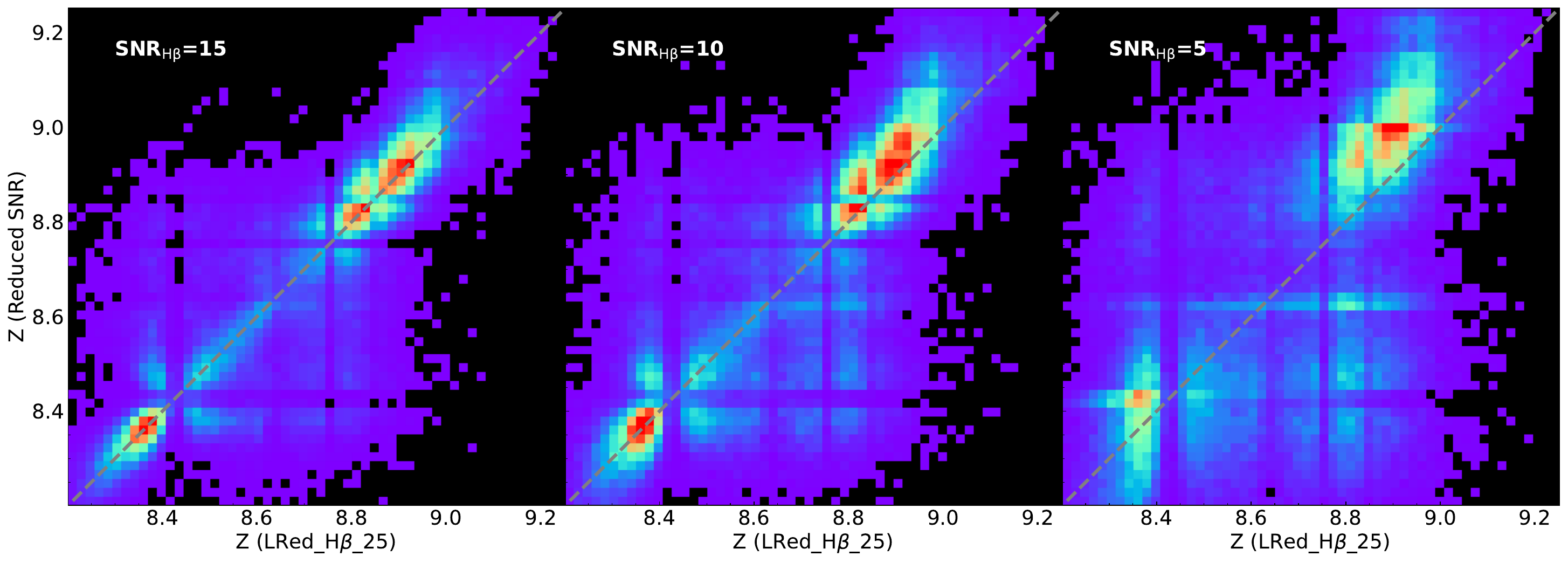}}
    \caption{The SNR tests for \texttt{LRed\_O3N2\_S2} combination in LRed suite for BPT-SF galaxies. The reference SNR$_{\mathrm{H}\beta}=25$, and we test 15, 10, and 5 in three panels. The counts of each bin are normalised to the maximum of each panel. From purple to red, the histograms show the range from 0 to 0.9.}
    \label{fig:snr_sf_lred}
\end{figure*}

\begin{figure*}
    \resizebox{16cm}{!}{\includegraphics{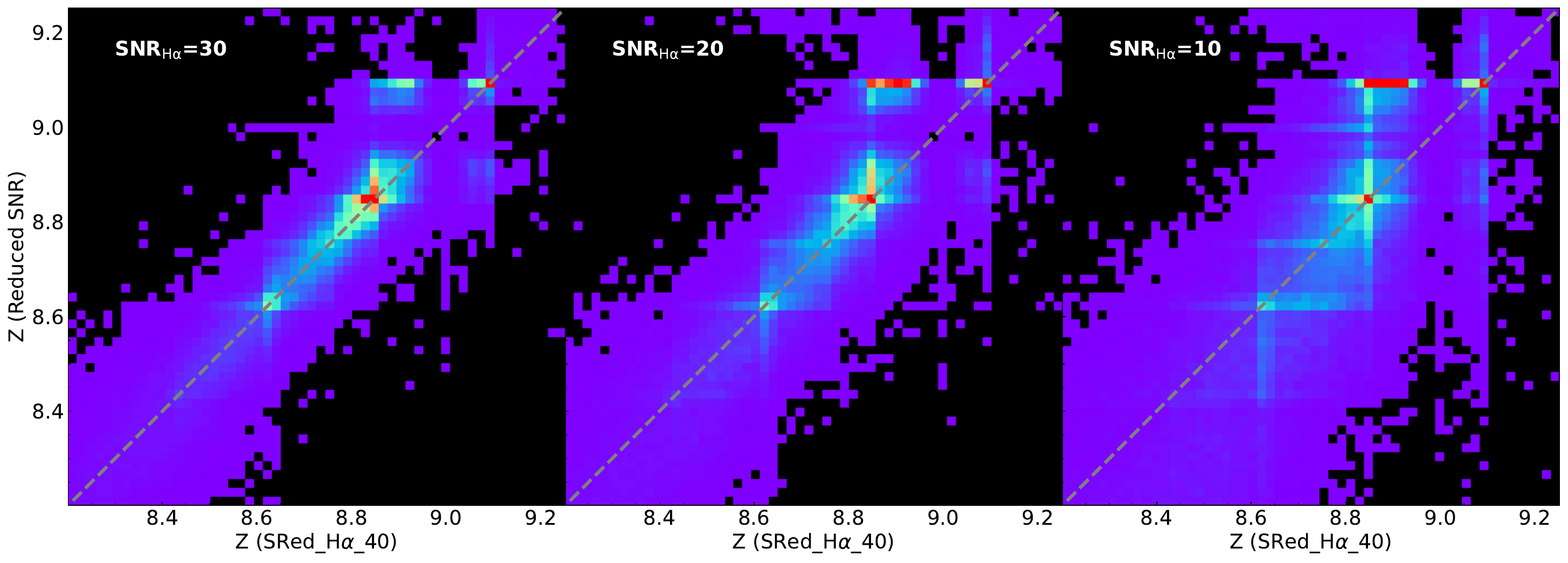}}
    \caption{The SNR tests for \texttt{SRed\_NS\_S2} combination in SRed suite for BPT-SF galaxies. The reference SNR$_{\mathrm{H}\alpha}=40$, and we test 30, 20, and 10 in three panels. The counts of each bin are normalised to the maximum of each panel. From purple to red, the histograms show the range from 0 to 0.5.}
    \label{fig:snr_sf_sred}
\end{figure*}

\begin{figure*}
    \resizebox{16cm}{!}{\includegraphics{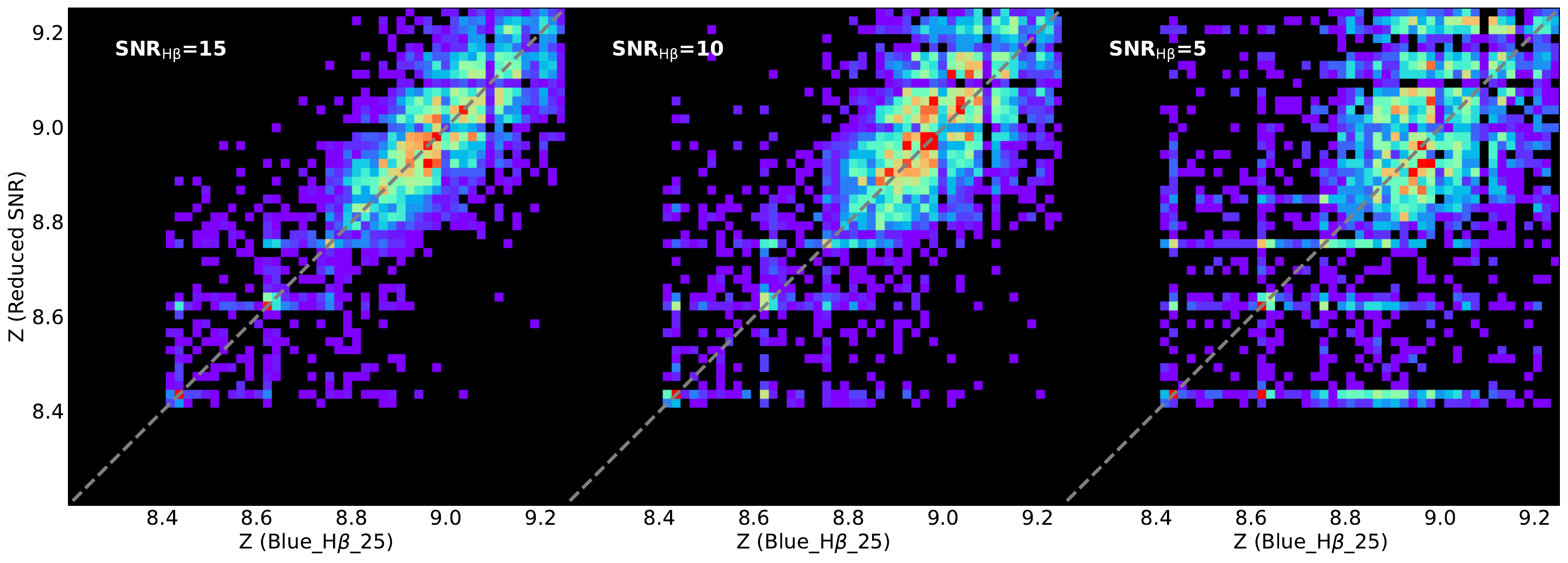}}
    \caption{The SNR tests for \texttt{Blue} combination in Blue suite for AGN-host galaxies. The reference SNR$_{\mathrm{H}\beta}=25$, and we test 15, 10, and 5 in three panels. The counts of each bin are normalised to the maximum of each panel. From purple to red, the histograms show the range from 0 to 0.9.}
    \label{fig:snr_agn_blue}
\end{figure*}

\begin{figure*}
    \resizebox{16cm}{!}{\includegraphics{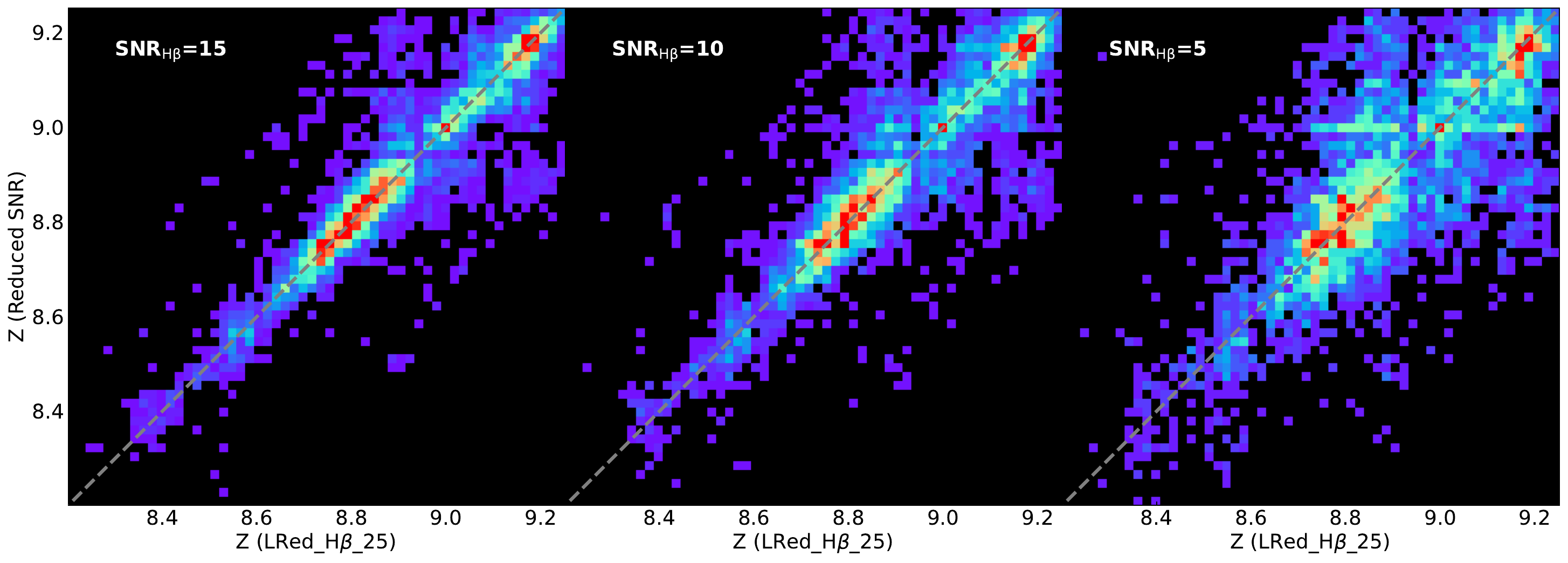}}
    \caption{The SNR tests for \texttt{LRed\_NS\_S2} combination in LRed suite for AGN-host galaxies. The reference SNR$_{\mathrm{H}\beta}=25$, and we test 15, 10, and 5 in three panels. The counts of each bin are normalised to the maximum of each panel. From purple to red, the histograms show the range from 0 to 0.6.}
    \label{fig:snr_agn_lred}
\end{figure*}

\begin{figure*}
    \resizebox{16cm}{!}{\includegraphics{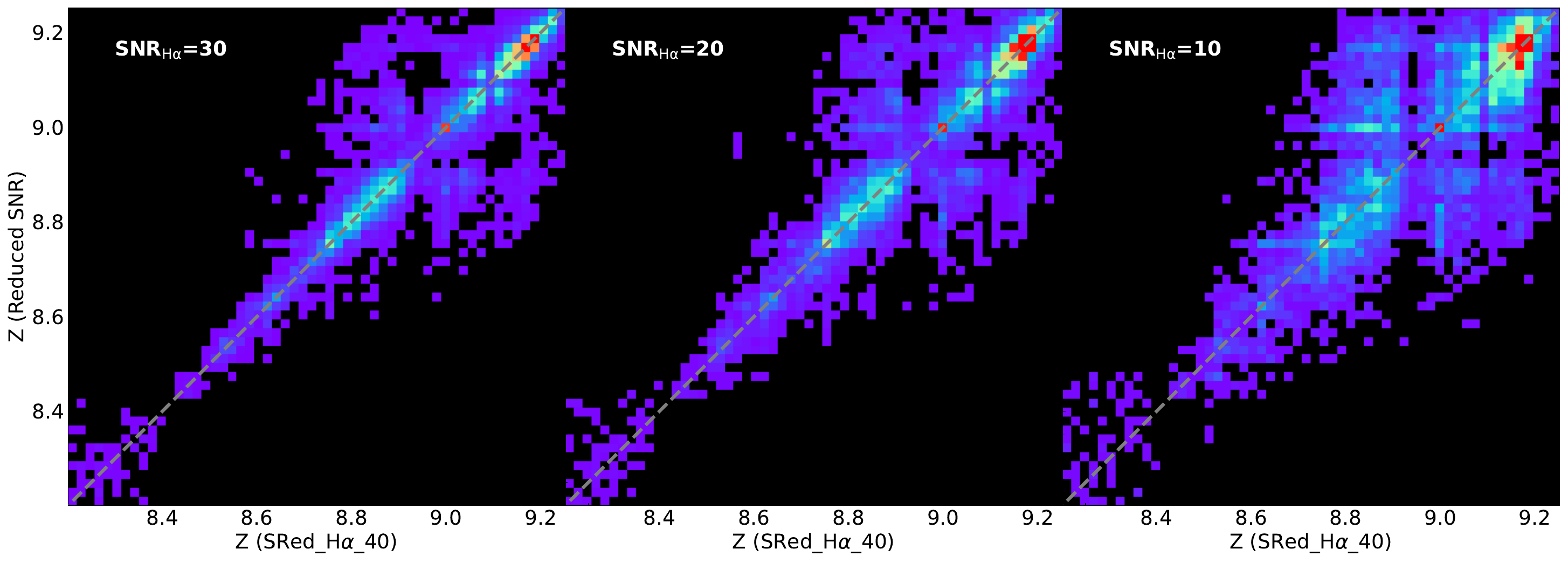}}
    \caption{The SNR tests for \texttt{SRed} combination in SRed suite for AGN-host galaxies. The reference SNR$_{\mathrm{H}\alpha}=40$, and we test 30, 20, and 10 in three panels. The counts of each bin are normalised to the maximum of each panel. From purple to red, the histograms show the range from 0 to 0.5.}
    \label{fig:snr_agn_sred}
\end{figure*}

In this appendix, we show all the SNR test plots corresponding to \autoref{sec:snr}, and the the statistics in \autoref{tab:snr_sf} and in \autoref{tab:snr_agn} are based on these plots. \autoref{fig:snr_sf_blue}, \autoref{fig:snr_sf_lred}, and \autoref{fig:snr_sf_sred} show the SNR tests for the best combinations in Blue, LRed, and SRed suites in BPT-SF galaxies, respectively. \autoref{fig:snr_agn_blue}, \autoref{fig:snr_agn_lred}, and \autoref{fig:snr_agn_sred} show the SNR tests for the best combinations in Blue, LRed, and SRed suites in AGN-host galaxies, respectively.


\bsp	
\label{lastpage}
\end{document}